\documentclass[twocolumn]{aastex6}

\usepackage{amsmath,amssymb,gensymb,times,graphics,morefloats,color}
\usepackage{afterpage, bm}
\usepackage{natbib,hyperref}
\def\kepler{\emph{Kepler}}
\def\ktwo{\emph{K2}}

\newcommand{\msun}{\rm M_{\sun}}
\newcommand{\rsun}{\rm R_{\sun}}

\def\vsini{$v\sin{i}$}

\def\numrotators{234}
\def\numyes{142}
\def\numyesbut{92}
\def\numpossibly{47}
\def\numall{574}

\definecolor{orange}{rgb}{.8,0.4,0}
\definecolor{blue}{rgb}{.0,0.0,1}

\usepackage{soul}

\shorttitle{ }
\shortauthors{Newton et al.}

\begin{document}

\title{New rotation period measurements for M dwarfs in the southern hemisphere: an abundance of slowly rotating, fully convective stars}

\author{Elisabeth R. Newton\altaffilmark{1,2}, Nicholas Mondrik\altaffilmark{3,4,5}, Jonathan Irwin\altaffilmark{5}, Jennifer G. Winters\altaffilmark{5}, David Charbonneau\altaffilmark{5} \\
}
\altaffiltext{1}{Massachusetts Institute of Technology Kavli Institute for Astrophysics and Space Research, 77 Massachusetts Avenue, Cambridge, MA 02139, USA}
\altaffiltext{2}{NSF Astronomy and Astrophysics Postdoctoral Fellow}
\altaffiltext{3}{Department of Physics, Harvard University, 17 Oxford St., Cambridge, MA 02138, USA}
\altaffiltext{4}{LSSTC Data Science Fellow, NSF Graduate Research Fellow}
\altaffiltext{5}{Harvard-Smithsonian Center for Astrophysics, 60 Garden Street, Cambridge, MA 02138, USA}

\begin{abstract}

Stellar rotation periods are valuable both for constraining models of angular momentum loss and for understanding how magnetic features impact inferences of exoplanet parameters. Building on our previous work in the northern hemisphere, we have used long-term, ground-based photometric monitoring from the MEarth Observatory to measure $\numrotators$ rotation periods for nearby, southern hemisphere M dwarfs. Notable examples include the exoplanet hosts GJ 1132, LHS 1140, and Proxima Centauri. We find excellent agreement between our data and \ktwo\ photometry for the overlapping subset. Amongst the sample of stars with the highest quality datasets, we recover periods in $66\%$; as the length of the dataset increases, our recovery rate approaches 100\%. The longest rotation periods we detect are around $140$ days, which we suggest represent the periods that are reached when M dwarfs are as old as the local thick disk (about $9$ Gyr).

\end{abstract}

\section{Introduction}

\subsection{Measuring rotation periods to inform studies of angular momentum evolution}

Stars are born with rotation periods between one and ten days, and experience initial spin-up as they contract onto the main sequence. Stars with masses below the Kraft break \citep{Schatzman1962, Kraft1967}, that is, which have convective envelopes, subsequently lose angular momentum through the coupling of the stellar wind to the magnetic field. The distribution of rotation periods has been well explored for solar-type stars from both an observational and a theoretical perspective \citep[see][for a review]{Bouvier2013}. 
Results from clusters at ages up to $4$ Gyr \citep{Barnes2016} indicate agreement between the observed stellar rotation periods and previously developed rotation--age  \citep[``gyrochronology'';][]{Barnes2003} relations.

Long-lived M dwarfs provide the opportunity to test the angular momentum evolution models that are traditionally tuned to the Sun. Models parameterize the way in which the magnetic field and mass loss produce a torque, which acts to remove angular momentum from the star; the magnetic field strength and topology and the mass-loss rate are thought to depend on the star's mass and rotation rate. These models are fit to the ages and rotation rates of groups of stars, with most models to date having focused on partially convective stars similar to the Sun. Several works have recently assessed models for both partially and fully convective stars, but there remain challenges in fitting both groups and in particular the slowly rotating fully convective stars \citep[e.g.][]{Reiners2012b,Matt2015,Douglas2016}.

The lowest mass M dwarfs are additionally challenging for theoretical models because much of their angular momentum evolution takes places at ages older than the oldest clusters in which such stars are accessible. M dwarfs in the field of the galaxy therefore provide important probes of spin-down. Photometric surveys over the past ten years, many of which have been motivated by the search for exoplanet transits, have yielded a plethora of rotation periods for field M dwarfs. Significant samples of rotating field M dwarfs have been contributed by HAT-Net \citep{Hartman2011}, MEarth \citep{Irwin2011, Newton2016}, Pan-STARRS \citep{Kado-Fong2016}, and \kepler\ \citep{McQuillan2013}. These studies of field stars complement those undertaken for clusters from the ground \citep[e.g.][]{Irwin2007, Hartman2010, Covey2016} and from space \citep[e.g.][]{Douglas2017, Rebull2016a}.

Mid-to-late M dwarfs ($0.08\lesssim M_* \lesssim 0.3\msun$) have not generally been accessible in wide-field surveys due to their intrinsic faintness. Thus, in order to connect the rotational evolution of these very low mass stars to F, G, K, and early M dwarfs, we must independently seek out their rotation periods.

\subsection{The importance of rotation periods to exoplanet research}

Not only are the rotation periods of field stars critical for stellar physics, but they are also important for enabling exoplanet research. Stellar rotation imprints itself on both photometry and radial velocity data, and thus the detection and characterization of planets near the stellar rotation period or its low-order harmonics can be frustrated. For early M dwarfs, the typical rotation periods for older field stars coincides with orbital periods of planets in the habitable zone \citep[]{Newton2016a, Vanderburg2016}. This can inhibit the detection of potentially habitable planets around these stars \citep[e.g. Gl 581;][]{Robertson2014} or the determination of masses for temperate planets discovered via their transits \citep[e.g. K2-3;][]{Almenara2015, Damasso2018}. 

The Transiting Exoplanet Survey Satellite \citep[\emph{TESS};][]{Ricker2014} launched in April 2018 with the goals of finding small planets around bright, nearby stars and measuring the masses for $50$ planets with radii $<4 R_\earth$. \emph{TESS}'s observing window ranges from $27$ days near the ecliptic to one year at the ecliptic poles, with one year spent in each of the ecliptic hemispheres over the course of its nominal two-year mission.  

\emph{TESS}'s time baseline combined with its small aperture means that the habitable-zone planets to which \emph{TESS} will be sensitive are those that orbit M dwarfs \citep[see Fig.~21 in ][]{Sullivan2015}. Though it will be possible to identify some periods from the \emph{TESS} data alone, the typically short time baseline will inhibit the detection of the long rotation periods common among M dwarfs. Thus, ancillary data will be needed to make the best use of limited resources for follow-up. Obtaining an optical spectrum to measure stellar magnetic activity provides one easily accessible avenue. For example \citet{Astudillo-Defru2017} calibrated period as a function of Ca H\&K emission strength. In \citet{Newton2017} we noted that if an M dwarf is inactive at H$\alpha$ it definitively identifies it as a slow rotator and presented a mass-period relation for H$\alpha$-inactive M dwarfs. 

Measuring the photometric behavior of a particular planet-host of interest is important for constraining the impact of spots on transit properties, and for modeling magnetic activity signatures in radial velocity measurements. Slowly rotating, inactive M dwarfs may represent more hospitable environments for life. This is because high-energy radiation and stellar mass loss can result in extensive atmospheric erosion for temperate planets around active M dwarfs \citep[e.g.][]{Lammer2007, Luger2015, Garraffo2017, Cohen2018}. Slowly rotating M dwarfs are likely to have fewer flares and CMEs, weaker stellar winds, and lower levels of x-ray and UV emission. Ultimately, it may also be possible to use the rotation period of an M dwarf to determine its age via gyrochronology \citep{Barnes2003}.

\subsection{This work}

The photospheres of distant stars are thought to be blemished by persistent magnetic features akin to the spots seen on the Sun.  The presence of spots modulates the stellar brightness as they rotate in to and out of view; measuring the periodicity of brightness variations therefore allows one to infer the rotation period of the star. The result is typically referred to as a photometric rotation period, which contrasts with periods inferred from velocity broadening of spectral lines ($v\sin{i}$).

Our work in \citet{Newton2016} provides the majority of photometric rotation period measurements for field M dwarfs below the fully convective boundary. These 387 rotation period measurements are based on photometry from the MEarth Project, using data obtained with the project's northern observatory, MEarth-North,  located at the Fred Lawrence Whipple Observatory, on Mount Hopkins, Arizona.  In this paper, we extend our analysis to the southern sky using data from MEarth-South. As for our northern search, our current work benefits from our survey's multi-year time baseline and consistent observing setup.  \emph{TESS} will begin its survey in the southern ecliptic hemisphere, which overlaps substantially with our survey area; due to their proximity to Earth, many of the nearby southern M dwarfs that are the subject of this work will be prime targets during \emph{TESS}'s first year for the detection of small planets and characterization of their masses and atmospheres.

\section{Our southern M dwarf sample}\label{Sec:data}

\subsection{The MEarth Project}

The MEarth Project is an all-sky survey of approximately 3000 nearby, predominantly mid-to-late M dwarfs \citep{Berta2012, Irwin2015}. The survey's second site, MEarth-South, was commissioned in January 2014. The survey design, data analysis, and period detection for MEarth-South are similar to those for MEarth-North. We provide an overview and point out differences here, but refer to \citet{Irwin2015} and \citet{Newton2016} for details. 

Like MEarth-North, MEarth-South comprises eight 40cm telescopes on German Equatorial Mounts, equipped with CCD cameras. These telescopes are housed in a roll-off roof at Cerro Tololo Inter-American Observatory (CTIO) in Chile. Exposure times are set independently for each star based on the S/N required to detect small planets. We use a maximum exposure time of $75$s, so this S/N is achieved by co-adding individual exposures. Multiple exposures are also obtained for bright targets in order to average over scintillation noise. We term one set of co-added exposures a \emph{visit}. There are typically between $1$ and $25$ visits per night, per target.

The MEarth-South survey is ongoing; the analysis presented here uses data obtained prior to BJD 2458179.5 (12am March 2 2018 UT). We analyze data from 574 stars from MEarth-South. Data from MEarth-North are not analyzed, except in \S\ref{Sec:k2} for stars that overlap with \emph{K2}. 
 
\subsection{Stellar parameters}

Trigonometric, photometric, and spectroscopic information were collected from the literature for stars in the MEarth target list. Table \ref{Tab:results} includes parallax, proper motion, radial velocity, and rotation period measurements, as well as several derived parameters. Parallaxes are not used when the measurement was less than three times the error in the measurement. When parallaxes are not available, we used photometric relations from \citet{Henry2004} to estimate distances; if optical (V, Rc, Ic) photometry is not available, estimates are taken from \citet{Lepine2005}.

We used 2MASS $K$ magnitudes to infer stellar masses and  applied the the mass--radius relation from \citet{Boyajian2012} to estimate stellar radii.  There are two commonly used relations to infer mass from $K$ magnitude: \citet{Benedict2016} and \citet{Delfosse2000}. The two calibrations agree within $0.01$ $\msun$ for M dwarfs with masses from $0.09$ to $0.22$ $\msun$, but the disagreement is worse at higher stellar masses. The results presented in this paper use \citet{Delfosse2000}, extrapolated as described in \citet{Newton2016}. This is for consistency with our previous work and because a mass--radius calibration has not been published for use with \citet{Benedict2016}. 

\subsection{MEarth-South target list}

The targets for MEarth-North were selected from the \citet{Lepine2005a} Northern proper motion catalog using distance \citep{Lepine2005}, radius, and color cuts \citep{Nutzman2008}. For MEarth-South \citep{Irwin2015}, we began with nearby M dwarfs with measured trigonometric parallaxes from the Research Consortium on Nearby Stars (RECONS)\footnote{http://www.recons.org/publishedpi.2012.1016} published in \citet{Winters2015} and spectroscopic characterization from the Palomar/Michigan State University (PMSU) spectroscopic survey \citep[][]{Reid1995, Hawley1996}. We then added stars from the LSPM-South catalog (S.~L\'epine, private communication), from which we used color and reduced proper motion cuts from \citet{Lepine2011} to select nearby M dwarfs. 

We then vetted stars to remove known close binaries prior to their inclusion in our southern target list. Rapidly rotating stars and close binaries are continuously removed from the active target list as they are identified. Stars are also typically removed from observation when they have $\geq$350 nights of observations. Our southern sample includes fewer M dwarfs above our fiducial radius limit ($0.33\rsun$) and beyond our fiducial distance limit ($33$ pc) than does the northern catalog due to the wider availability of accurate distance measurements and removal of binaries prior to the start of the survey. 

\begin{figure}
\begin{centering}
\includegraphics[width=\linewidth]{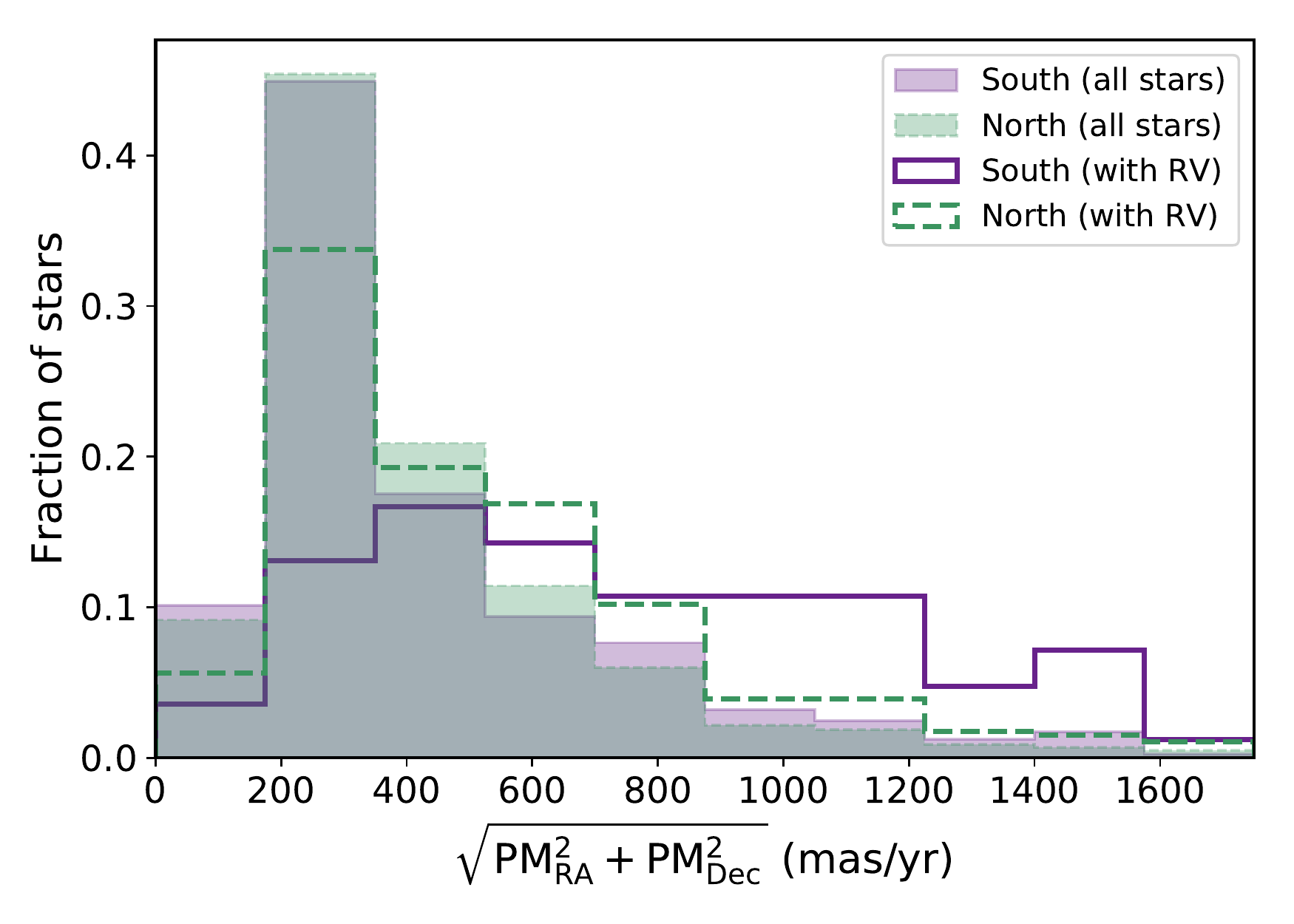}
\end{centering}
\caption{Comparison between the proper motions of the southern sample (purple) and the northern sample (teal). The filled histograms shows all stars; the unfilled histograms show only those with radial velocities. Due to the provenance of the southern radial velocities, the southern stars with all three components of motion measured are biased towards kinematically older stars.
\label{Fig:vspace-ns}}
\end{figure}

The census of M dwarfs within $33$ pc is not as complete in the South as it is in the North, which raises concern that the samples might differ in typical properties, such as age. Indeed, we note a significant bias towards kinematically older stars when examining the total space velocities of stars in each sample. However, the distribution of proper motions for stars in our northern and southern samples is similar, reflecting their common origin in the catalogs of high proper motion sources (Fig. \ref{Fig:vspace-ns}). 

We determine that the space velocity bias derives from the subset of each sample for which radial velocities are available. In the South, most radial velocity measurements available in the literature are from the PMSU Survey conducted in the 1990s, which targeted high proper motion stars included in the 3rd edition of the Catalog of Nearby Stars \citep[CNS3;][]{Gliese1991}. The CNS3 proper motion limits are higher than the limits of present-day proper motion surveys. Therefore, the southern stars with radial velocities are biased towards those with high proper motions. This effect is not seen in the north because numerous surveys have targeted northern M dwarfs, basing their target lists on more recent catalogs.

\vspace{10pt}

\section{Rotation period measurements}

\subsection{Period fitting algorithm}

To detect rotation periods, we use the ``least squares periodogram'' method as described in \citet{Irwin2006}. The application to MEarth data is described in \citet{Irwin2011} and \cite{Newton2016}. We simultaneously fit for systematics and a sinusoid. Our systematics includes magnitude offsets corresponding the changes in our detectors and for observations obtained on each side of the meridian, and the ``common mode''. The common mode derives from variations in precipitable water vapor, which differentially affects our late-type science targets and our earlier-type comparison stars; as will be discussed later, this impacts our ability to detect rotation periods. The sinusoid we presume to be the result of starspots rotating in and out of view. 

All parameters except the period are fit using linear least squares. We step through a grid of rotation periods and record the F-test goodness of fit metric at each period, the result of which is a periodogram where the power is given by the F-test statistic. The candidate period is the one with the highest F-test statistic. 

We initially search periods between $0.1$ and $1000$ days, but also explore parameter space manually using a custom GUI.\footnote{github.com/ernewton/rogui \label{rogui}} For stars with candidate periods $<0.2$ days, we repeated the search allowing periods as short as $0.05$ days. A typical occurrence is that longterm, non-repeating variability will manifest as either a $1$ day period or a very long ($500-1000$ day) period. We try searching narrower period ranges for these stars. The final recorded rotation period represents a decision made by a human and may not be the highest peak in the periodogram.

We treat each dataset obtained with a single telescope and single filter as one continuous timeseries and do not introduce any additional breaks. The period of the sinusoid is the only parameter fixed between different timeseries for one star.  Observations of many stars with MEarth-North consisted of two or more timeseries with splits occurring every one to four years primarily as a result of filter changes. In contrast, most MEarth-South stars have only one timeseries: they been consistently observed with one telescope, and the filters on MEarth-South have not been changed. This implicitly assumes that for, the purposes of period detection, we can model a multi-year timeseries as a single, unchanging sinusoid.

\afterpage{
\begin{deluxetable*}{l l l l} 
\tablecaption{\label{Tab:results}Rotation periods and stellar properties for all rotators and non-detections in MEarth-South (table format)}
\tablecolumns{4}
\tablewidth{6.4in}
\tablehead{\colhead{Column} & \colhead{Format} & \colhead{Units} & \colhead{Description}}
\startdata
   \vspace{-10pt} \\ 
   1 & A1  & $\cdots$ &     	\parbox{4in}{ Source type:    
    A = Grade A rotators;
    B = Grade B rotators;
    U = Possible or uncertain detection;
    N = Non-detection or undetermined detection.}
 \\
   \vspace{-10pt} \\ 
   2 & A17  & $\cdots$ &    2MASS identifier \\
  3 & A11    & $\cdots$&    LSPM identifier  \\
  4 &  F10.6 &  deg &   	RA in decimal degrees (J2000)\\
  5 & F11.8  & deg   &    	Dec. in decimal degrees (J2000)\\
  6 & F7.4  &  arcsec/yr &   Proper motion along RA \\
  7 & F7.4  &  arcsec/yr & Proper motion along Dec. \\
   8 & F6.4  &  arcsec  & 	Parallax\\
  9 & F6.4   & arcsec  & 	Uncertainty in parallax \\
  10 &  A19    & $\cdots$ &   ADS bibliography code reference for parallax \\
  11 & F5.2   & pc  &           Adopted distance \\
 12 & F6.1  &  km/s      &    Radial velocity \\
 13 &  F4.1  &  km/s    &     Uncertainty in radial velocity \\
 14 &  A19    & $\cdots$  & ADS bibliography code reference for RV \\
  15 & F8.3   & days    &    	Photometric rotation period\\
 16 & F6.4  &  mag   &   	Semi-amplitude of variability\\
 17  & F6.4 &   mag  &     	Uncertainty in semi-amplitude\\
 18 & F5.3  &  $\msun$   & 	Stellar mass\\
 19 & F5.3  &  $\rsun$   &  	Stellar radius\\
 20 & I1   &   $\cdots$  & 	Flag indicating known contamination by a common 
          proper motion companion or background source\\
 21 & I5  &    $\cdots$    &  Number of data points in longest dataset\\
 22 & I4  &    $\cdots$    &  Number of days in longest dataset\\
 23  & F6.4   &  mag &  Median photometric error \\
 24  & I6    &     $\cdots$ &  F-test statistic  
\enddata
\end{deluxetable*}
\clearpage
}

\subsection{Classifying rotators, candidates, and non-detections}

We assess the validity of the candidate periods by eye and assign ratings as in \citet{Newton2016}. Stars for which we are confident that the variability is intrinsic to the star and for which a periodicity can definitively be assigned are considered to be rotation period detections. We classify rotators as ``grade A'' or ``grade B''. We further identify candidate rotators that receive ratings of ``possible/uncertain'', and classify the remainder as ``non-detection/undetermined.''  The specific questions that we pose in order to make these assignations are described in \citet{Newton2016}.

Though comparison between our measurements and literature results suggests that we have identified the correct period for many candidate rotators, we limit our analysis to grade A and B rotators. We present our periods in Table \ref{Tab:results}; ratings are indicated in the first column by a letter. We do not report periods for non-detections, and note that periods for non-detections listed in \citet{Newton2016} should not be used.

Table \ref{Tab:results} contains periods for stars whose apertures are severely contaminated by a physically associated or background companion. These objects are flagged in the table but are not included in the analysis that follows.

\subsection{Comparison with K2}\label{Sec:k2}

\kepler\ observed some stars in our sample as part of the \ktwo\ ecliptic survey. We identified M dwarfs in the MEarth dataset that had also been observed by \ktwo. Several of these stars are part of the MEarth-North dataset and not in the sample of stars otherwise included in this work; these stars are analyzed in this section only. We downloaded the \ktwo\ lightcurves available on MAST as of January 17 2018. We considered every data reduction available on MAST but found that the K2SFF \citep{Vanderburg2014} and the PDC-MAP \citep{Stumpe2012, Smith2012} reductions best served our purposes. We use the \texttt{astropy} \citep{TheAstropyCollaboration2018} package to compute a Lomb Scargle periodogram and our custom visualization tool to examine each lightcurve by eye.\textsuperscript{\ref{rogui}} 

\begin{figure}
\begin{centering}
\includegraphics[width=\linewidth]{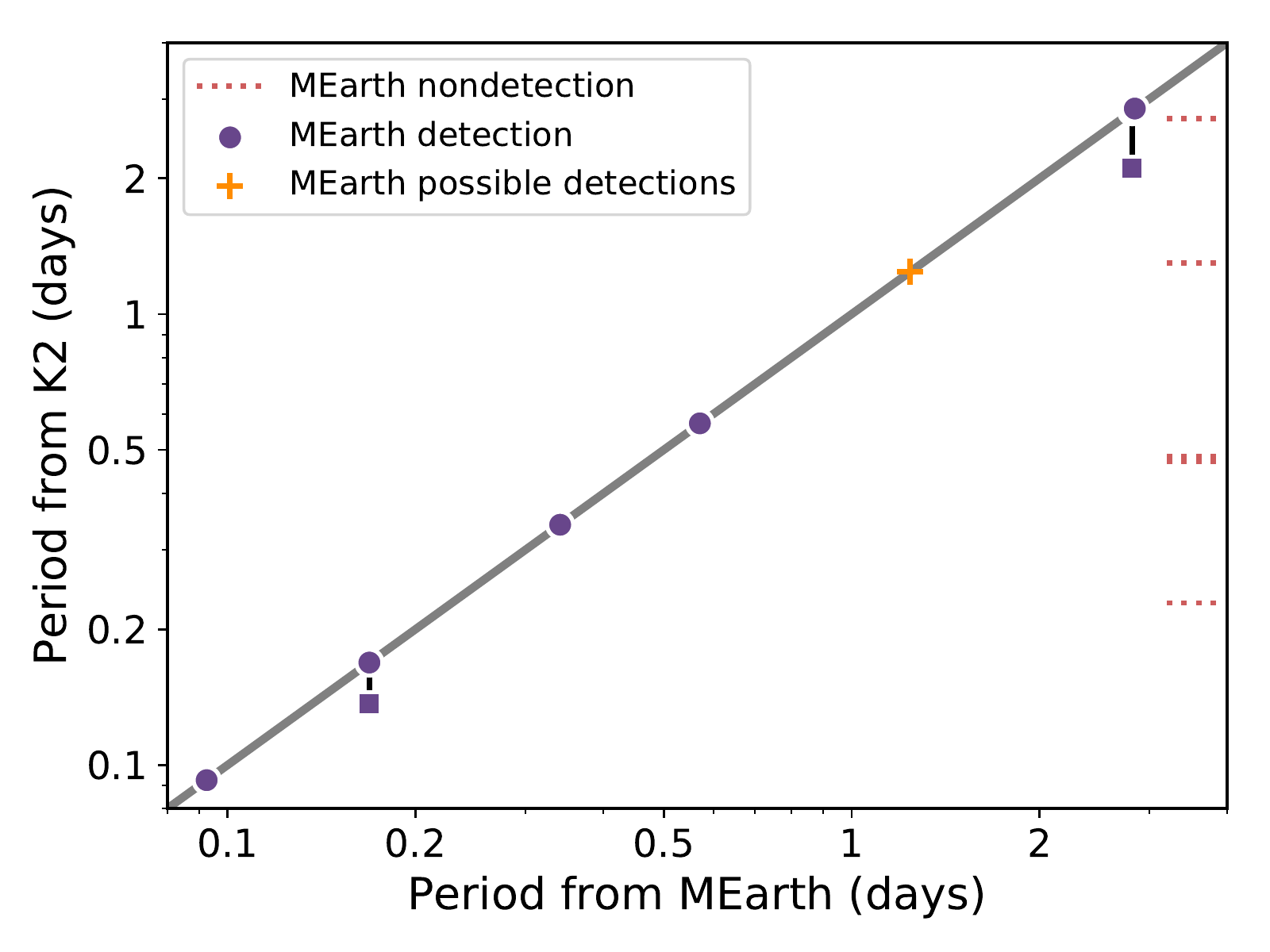}
\end{centering}
\caption{Comparison of rotation periods we measured using MEarth data to those we measured using \ktwo\ data. The solid line shows the one-to-one line. The two sets of vertically aligned points connected by a line represent cases where we clearly identified two rotation periods for a single target in \ktwo. The dotted lines along the right side of the figure indicate stars for which a period was identified from \ktwo\ but not from MEarth. Stars for which a period was identified from MEarth but not \ktwo\ are not shown; the stars that fall into this category have rotation periods longer than $50$ days. 
\label{Fig:k2}}
\end{figure}

We find clear rotation periods for $12$ stars from \ktwo, all of which are less than three days. Of these, seven have rotation periods from MEarth, all of which are in agreement. For 2MASS J11280702+0141304/EPIC 201577109 and 16204186-2005139/EPIC 204957517, the \ktwo\ lightcurves indicate two distinct rotation periods. This suggests that the targets may be blended, but could also indicate differential rotation. We present a comparison between the MEarth and \ktwo\ rotation periods in Tab. \ref{Tab:k2} and in Fig. \ref{Fig:k2}.

Five stars have rotation periods from \ktwo\ but not from our work with MEarth. For four of these, we have have $<100$ observations with MEarth and we do not expect to detect rotation with such limited data. The fifth star is Wolf 359 (2MASS J10562886+0700527, EPIC 201885041), an M6 dwarf \citep{Kirkpatrick1991} for which the \ktwo\ data clearly indicate a $2.7$ rotation period.\footnote{Though this star has been known to be strongly magnetically active, its rotation period has not been previously identified and no \vsini\ broadening is seen \citep[$v\sin{i}<2.5$ km/s;][]{Browning2010}. We note that given the star's small size, this is still consistent with being viewed edge-on, assuming $R_*=0.13$ $\rsun$.} This objects suffers from unfavorable observing conditions (see also \S\ref{Sec:nondetections}).

For the stars for which we have identified long rotation periods, we do not expect to detect rotation periods in the \ktwo\ data given the short length of the \ktwo\ campaigns, and indeed we find no periods for these long rotators. We jointly examine the MEarth and \ktwo\ lightcurves for each M dwarf with a confirmed or candidate rotation period from MEarth. Though the available data reductions do not aim to preserve stellar variability on the timescale of a \ktwo\ campaign, we find that the long-term modulations present in the K2SFF reduction are roughly consistent with what we would expect given the rotation period from MEarth.

\begin{figure*}
\centering
\includegraphics[width=0.85\linewidth]{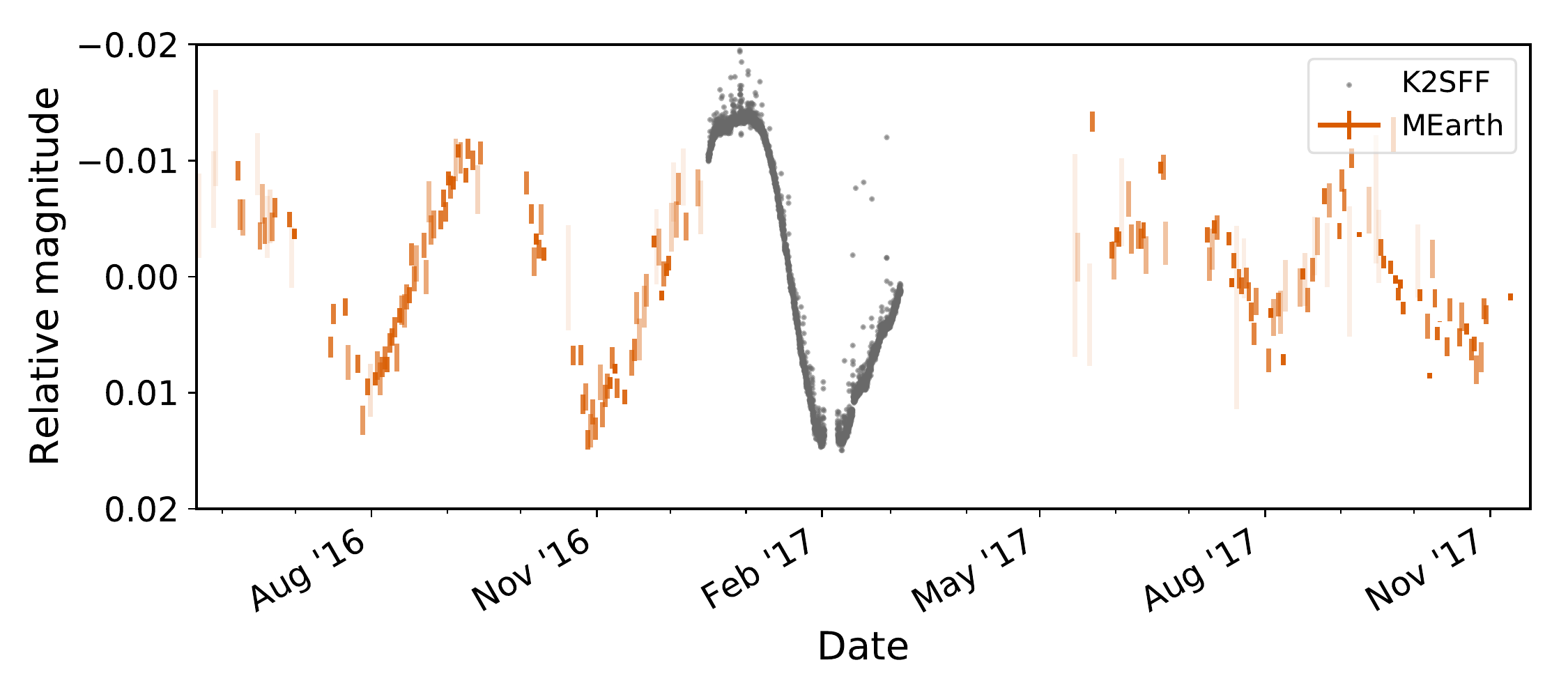}
\caption{Lightcurve for GJ 1286, showing both the MEarth (orange) and \emph{K2} (gray) datasets. The MEarth dataset has been median combined into one day bins, with the error bars representing an estimator for the sample standard deviation (1.48 times the median absolute deviation). The \emph{K2} data are from lightcurves produced by \citet{Vanderburg2014} and have been manually offset in magnitude to match our MEarth data. The \ktwo\ data are consistent with the $89$ day rotation period we detected in MEarth.
\label{Fig:gj1286}}
\end{figure*}

This is particularly striking for GJ 1286 (2MASS 23351050-0223214, EPIC 246333864), for which we detect rotational modulation at $89$ days. GJ 1286 is an M5.5 dwarf \citep{Henry1994} at 7.2 pc \citep[]{Weinberger2016}.  \emph{K2} Campaign 12 observations commenced as GJ 1286 set for our Earth-bound observatories. Comparing the MEarth and K2SFF lightcurves, it is clear that the two datasets are consistent, both showing near-sinusoidal modulation (Fig. \ref{Fig:gj1286}). 

\begin{deluxetable*}{l l r r l r r r}
  \tablecaption{ Stars in our sample also in \ktwo
  \label{Tab:k2} }
\tablehead{
\colhead{\textbf{2MASS ID}} &
\colhead{\textbf{EPIC ID} }&
\colhead{$P_{1}$ (K2) } &
\colhead{$P_{2}$\tablenotemark{a} (K2)} &
\colhead{\textbf{Source}\tablenotemark{b}} &
\colhead{$P$ (MEarth)} &
\colhead{\textbf{Rating}} &
\colhead{\textbf{$N_\mathrm{obs}$} }}
\startdata
10252645+0512391 & 248574998 & 0.09254 & $\ldots$ & k2sff & 0.09253 & A & 1124 \\
10562886+0700527 & 201885041 & 2.713 & $\ldots$ & ktwo & $\ldots$ & N & 2774 \\
11280702+0141304 & 201577109 & 0.1687 & 0.1369 & ktwo & 0.1687 & A & 1029 \\
12115719-0720136 & 228786554 & $\ldots$ & $\ldots$ & k2sff & 140.5 & B & 3796 \\
12220398-0629123 & 228807026 & 1.24 & $\ldots$ & k2sff-c102 & 1.24 & U & 681 \\
12235208-0858432 & 228748748 & 0.471 & $\ldots$ & k2sff & $\ldots$ & N & 9 \\
12384731-0419168 & 228858734 & $\ldots$ & $\ldots$ & ktwo & $\ldots$ & N & 1620 \\
13215411-1424098 & 212422696 & 1.296 & $\ldots$ & ktwo & $\ldots$ & N & 26 \\
13300285-0842251 & 212681564 & $\ldots$ & $\ldots$ & k2sff & $\ldots$ & N & 917 \\
13515712-1758490 & 212285198 & $\ldots$ & $\ldots$ & k2sff & $\ldots$ & N & 29 \\
16131715-2538139 & 203585853 & $\ldots$ & $\ldots$ & k2sff & $\ldots$ & N & 350 \\
16204186-2005139 & 204957517 & 2.82 & 2.106 & k2sff & 2.814 & B & 1212 \\
16352464-2718533 & 203083752 & $\ldots$ & $\ldots$ & k2sff & 122.7 & A & 3815 \\
16563362-2046373 & 204806561 & 0.5724 & $\ldots$ & k2sff-c111 & 0.5711 & A & 1980 \\
18471674-1922202 & 218121888 & $\ldots$ & $\ldots$ & k2sff & 152.1 & B & 1893 \\
18494929-2350101 & 215632123 & 2.857 & $\ldots$ & ktwo & 2.843 & A & 1428 \\
19042185-2406157 & 215493021 & $\ldots$ & $\ldots$ & ktwo & $\ldots$ & N & 620 \\
19162323-2322324 & 215875814 & $\ldots$ & $\ldots$ & ktwo & $\ldots$ & N & 2265 \\
22134277-1741081 & 205913009 & $\ldots$ & $\ldots$ & k2sff & $\ldots$ & N & 4194 \\
22260112-1518128 & 205983882 & 0.2288 & $\ldots$ & ktwo & $\ldots$ & N & 22 \\
22285440-1325178 & 206050032 & $\ldots$ & $\ldots$ & ktwo & $\ldots$ & N & 4680 \\
22341112-1020135 & 206169145 & 0.4834 & $\ldots$ & ktwo & $\ldots$ & N & 102 \\
23172072-0236323 & 246322698 & 0.341 & $\ldots$ & ktwo & 0.341 & B & 1581 \\
23180785-0234475 & 246324216 & $\ldots$ & $\ldots$ & k2sff & 114.5 & U & 1581 \\
23351050-0223214 & 246333864 & $\ldots$ & $\ldots$ & k2sff & 88.92 & A & 3624 \\
23552591-0359000 & 246253313 & $\ldots$ & $\ldots$ & k2sff & 54.47 & A & 2584
\enddata
\tablenotetext{a}{Secondary peak in the periodogram when clearly present.}
\tablenotetext{b}{Source of \ktwo\ indicated using the MAST prefix of the datafile used. ``k2sff'' is from \citet{Vanderburg2014} and ``ktwo'' is the PDC-MAP reduction. For Campaigns 10 and 11, the PDC-MAP data is split into two lightcurves and we have indicated the segment used.}
\end{deluxetable*}

\begin{figure*}
\centering
\includegraphics[width=0.48\linewidth]{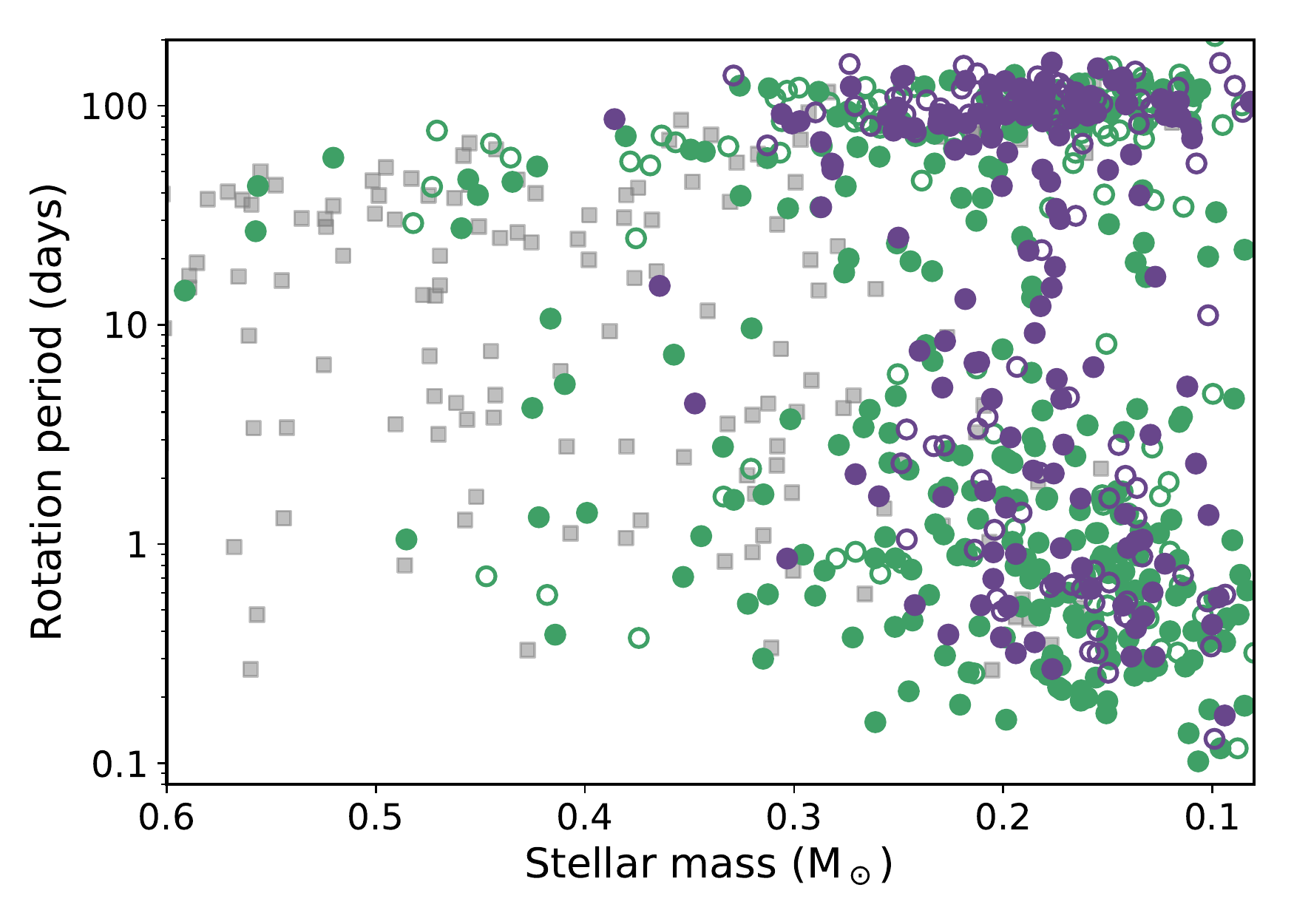}
\includegraphics[width=0.48\linewidth]{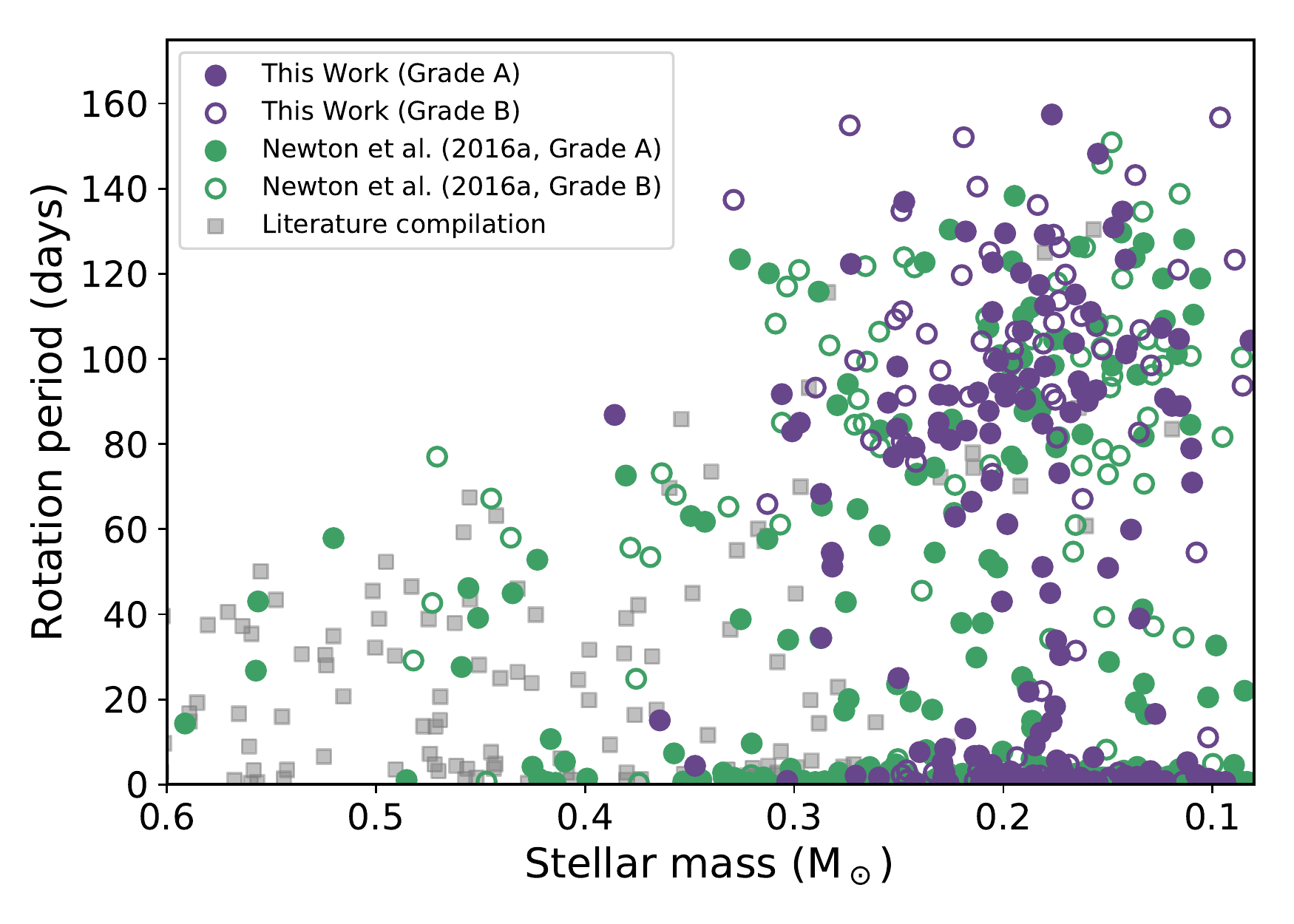}
\caption{Rotation period in days versus stellar mass for stars in the MEarth survey databases, with the vertical axes scaled logarithmically (left) and linearly (right). Gray squares are measurements compiled from the literature, many of which are from \citet{Hartman2011}. Green circles are measurements from \citet{Newton2016}, purple circles are measurements presented in this work. Filled circles are Grade A detections and open circles are Grade B detections; candidate rotators are not shown. 
\label{Fig:massper}}
\end{figure*}

\section{Results}

\subsection{Stars with detected rotation periods}

\begin{figure*}
\centering
\includegraphics[width=0.33\linewidth]{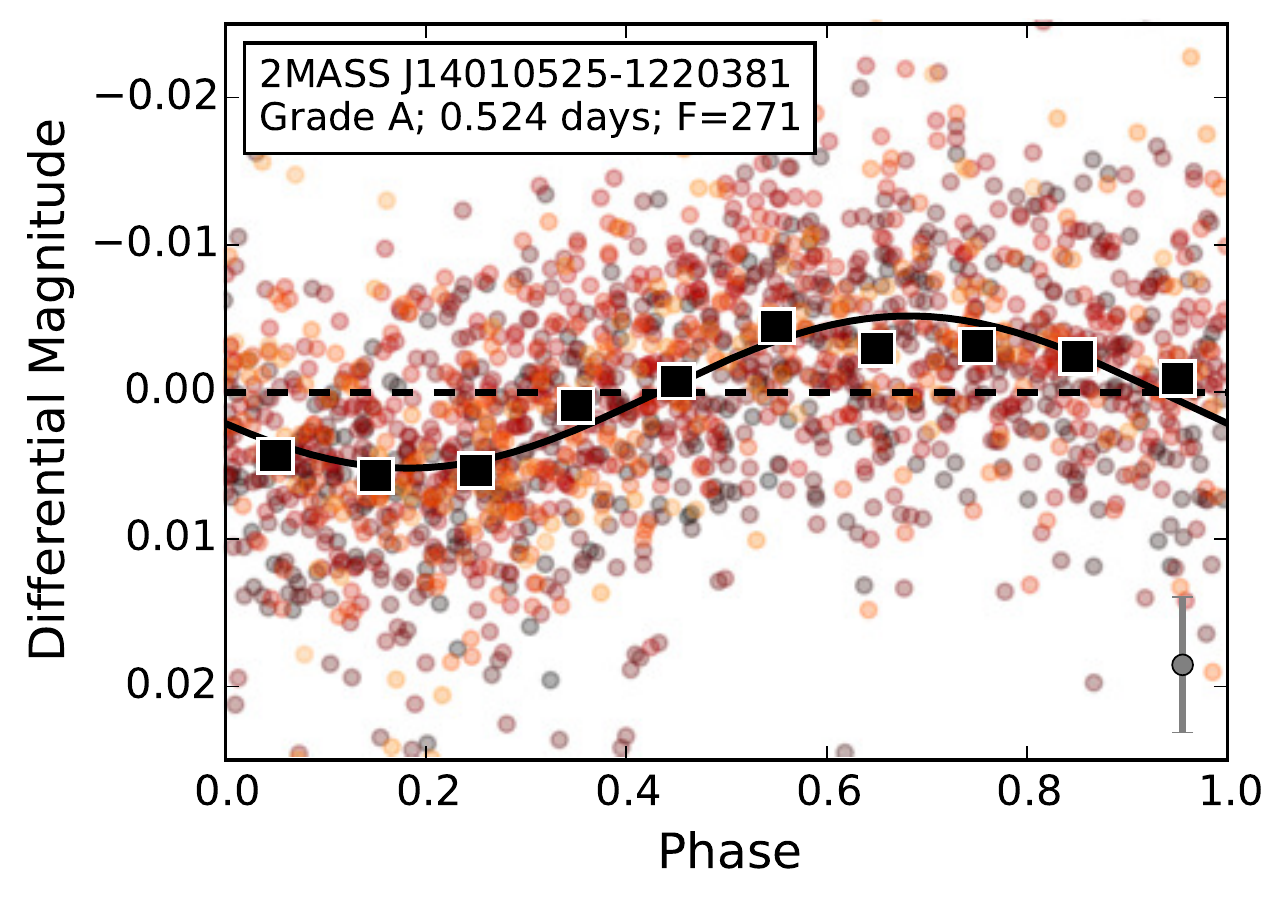}
\includegraphics[width=0.33\linewidth]{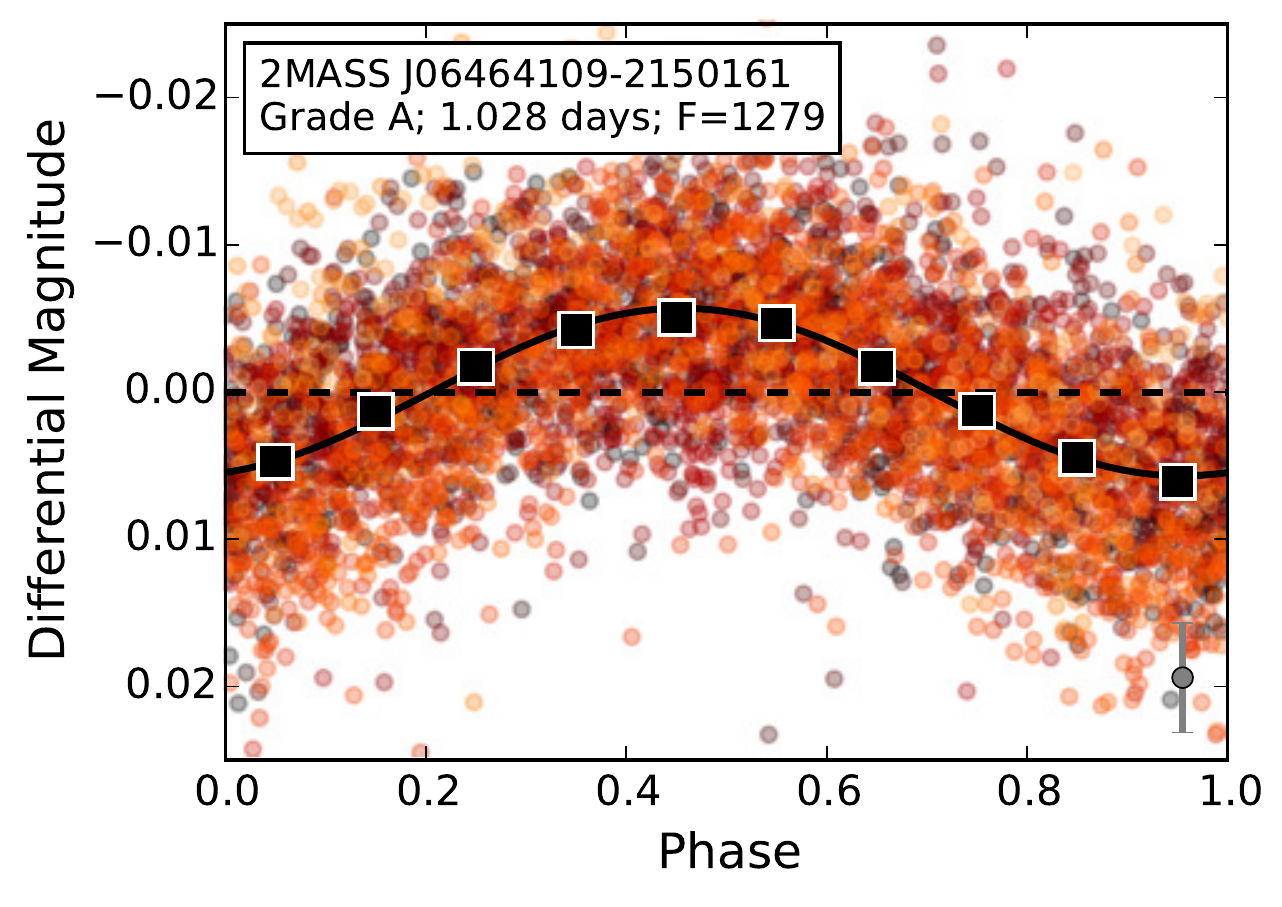}
\includegraphics[width=0.33\linewidth]{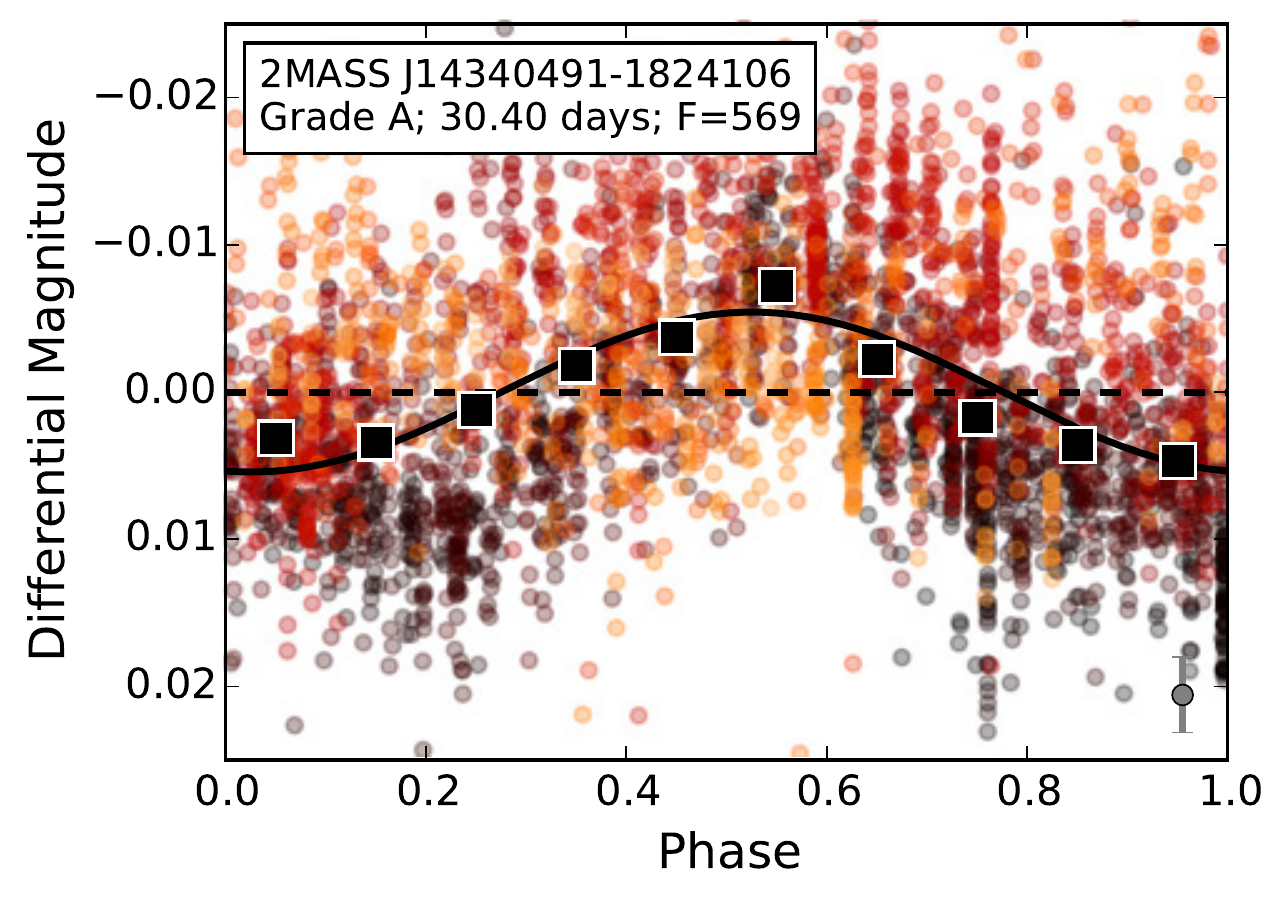} \\
\includegraphics[width=0.33\linewidth]{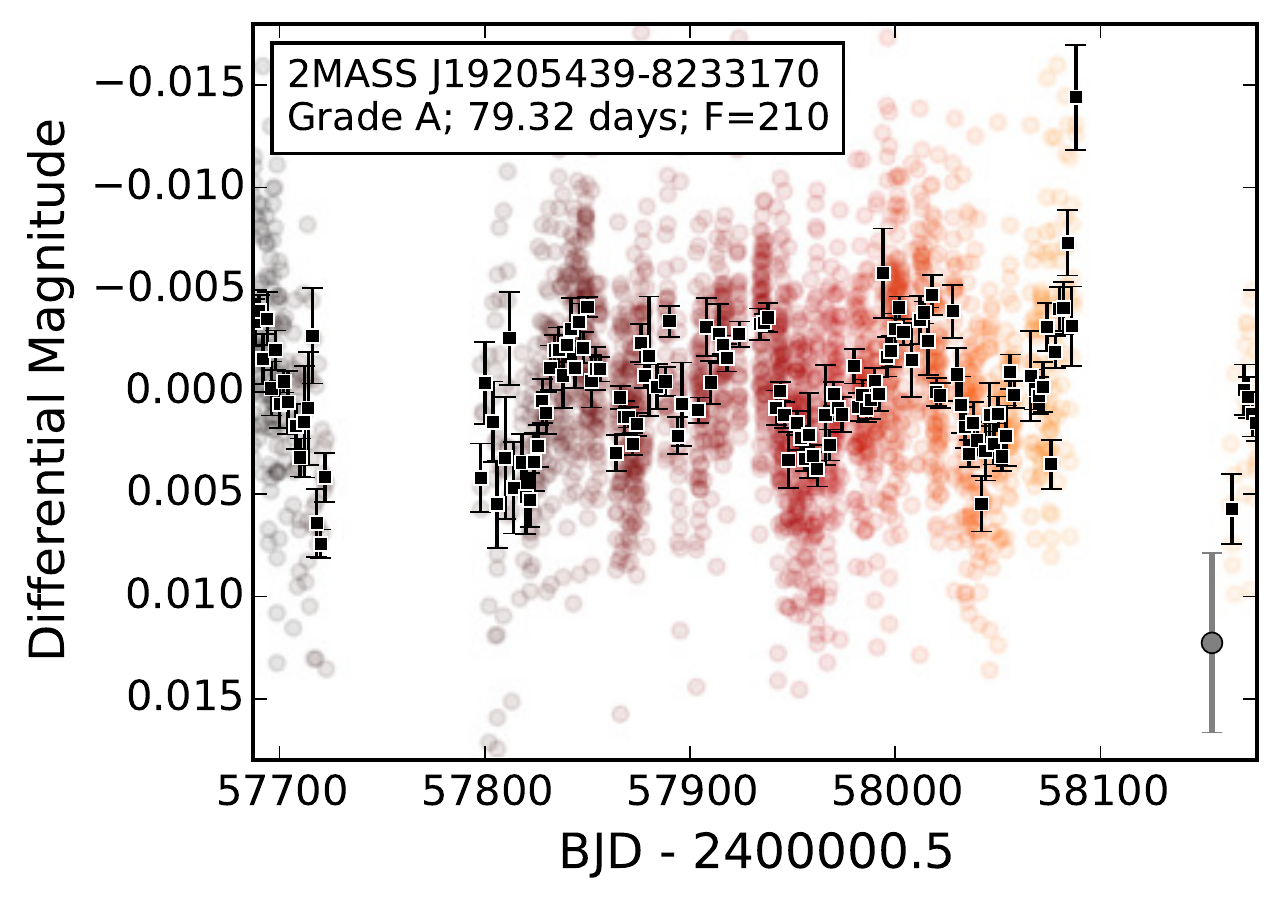}
\includegraphics[width=0.33\linewidth]{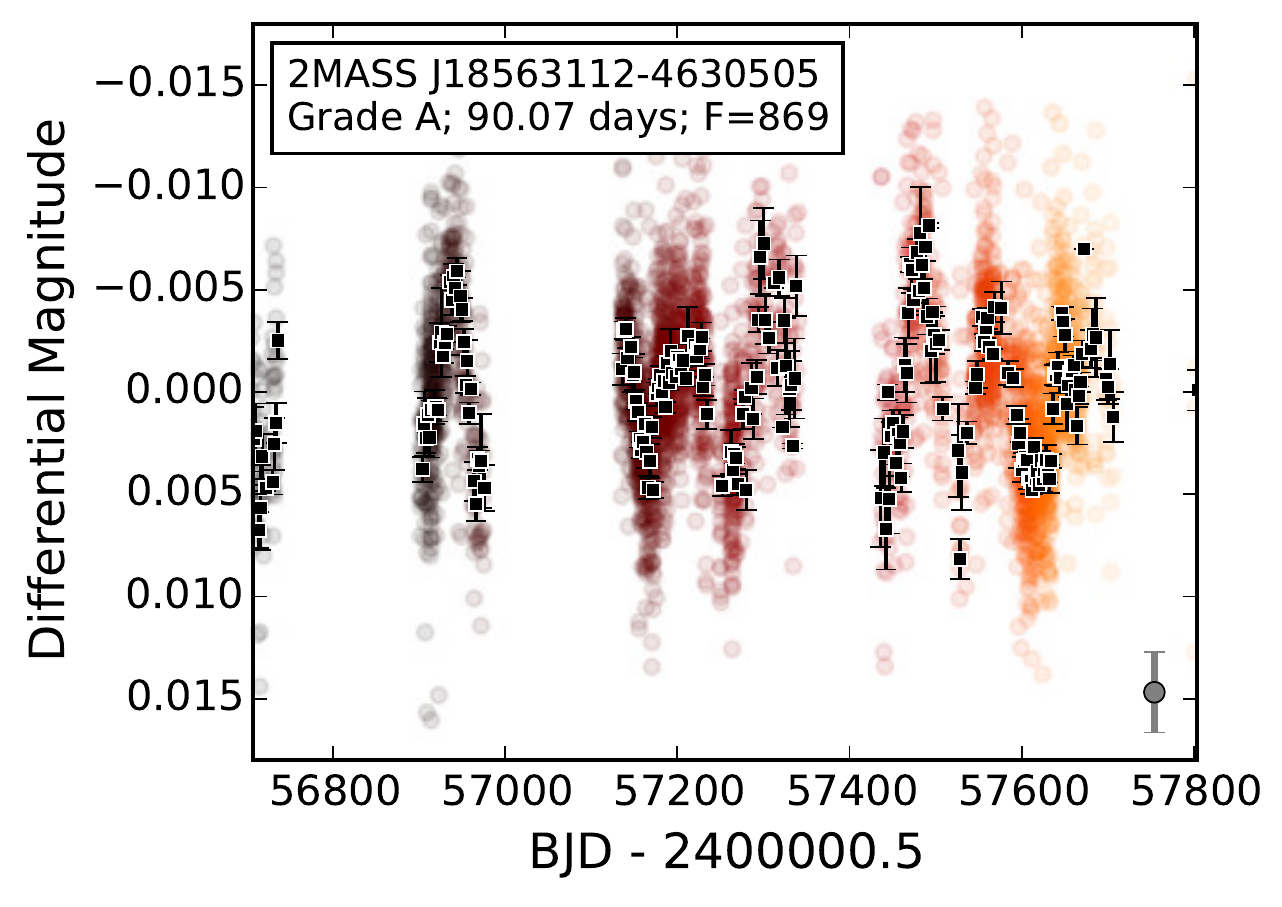}
\includegraphics[width=0.33\linewidth]{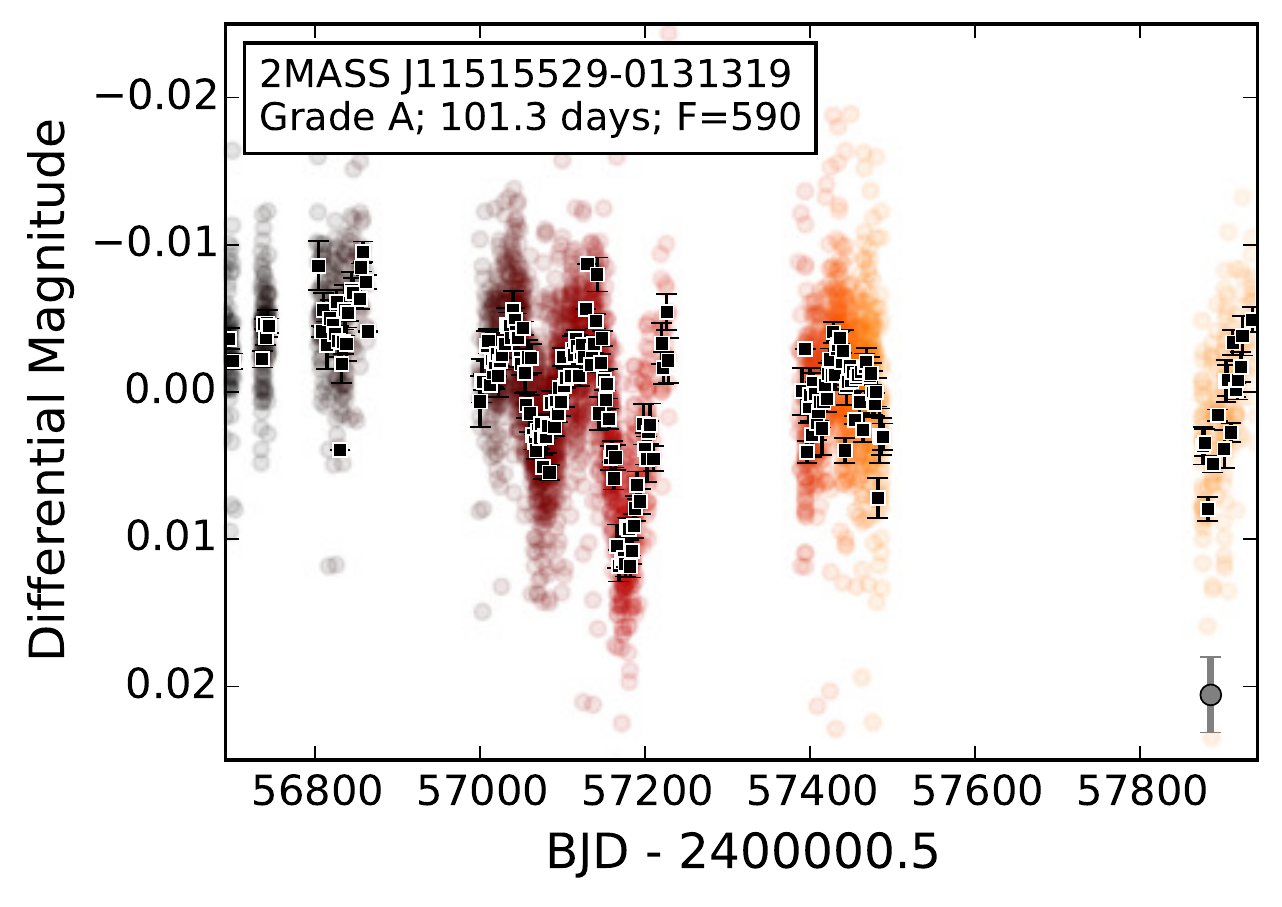}\\
\includegraphics[width=0.33\linewidth]{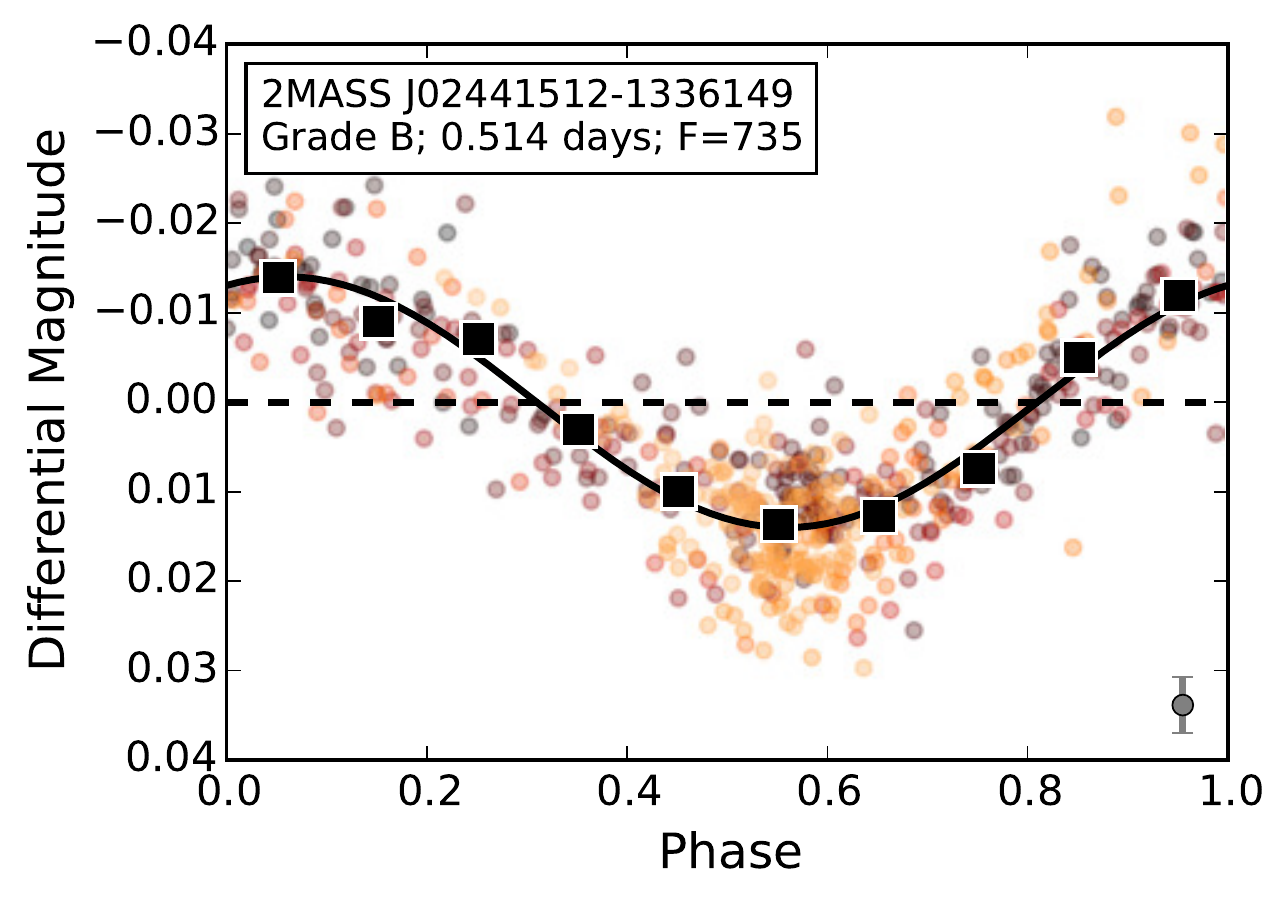}
\includegraphics[width=0.33\linewidth]{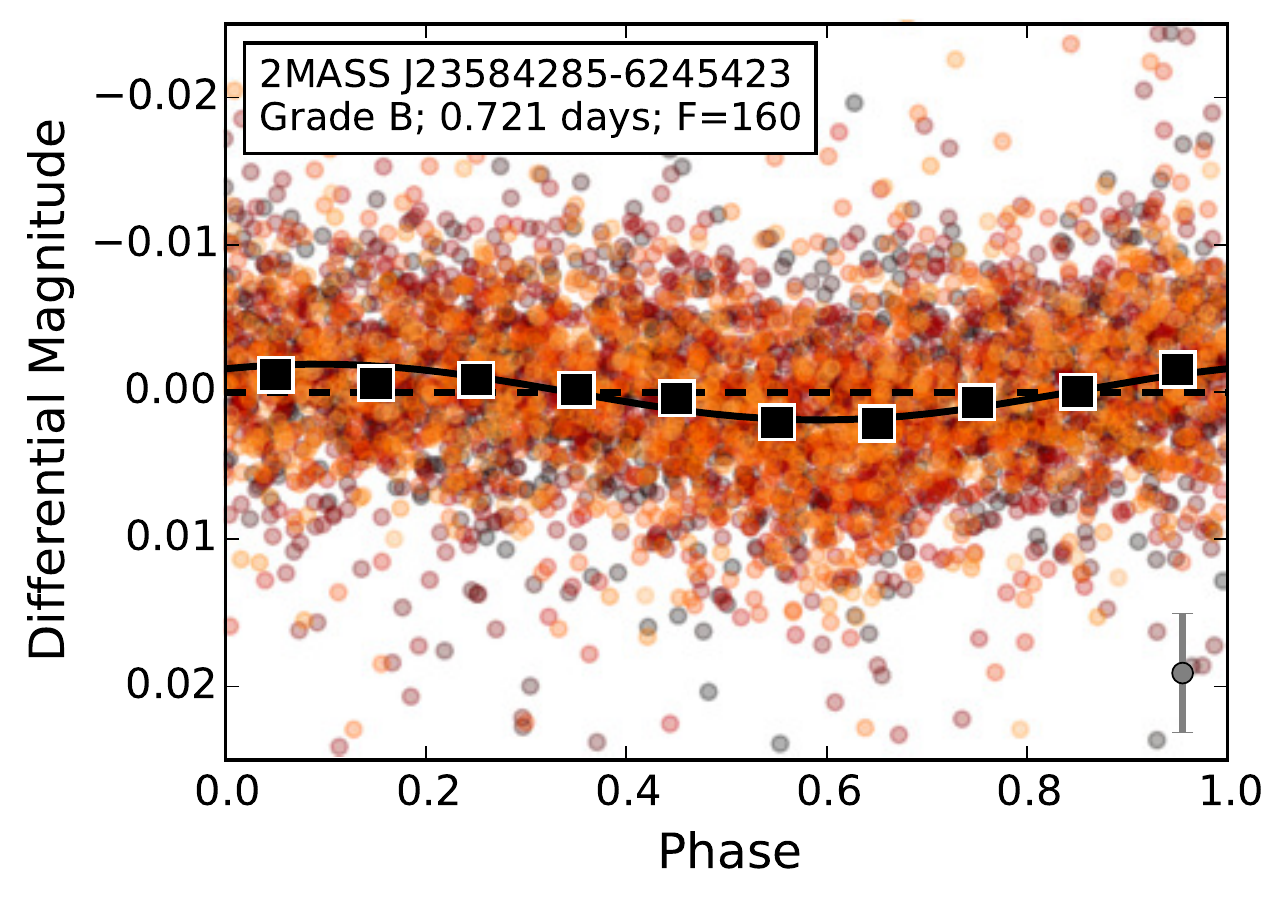}
\includegraphics[width=0.33\linewidth]{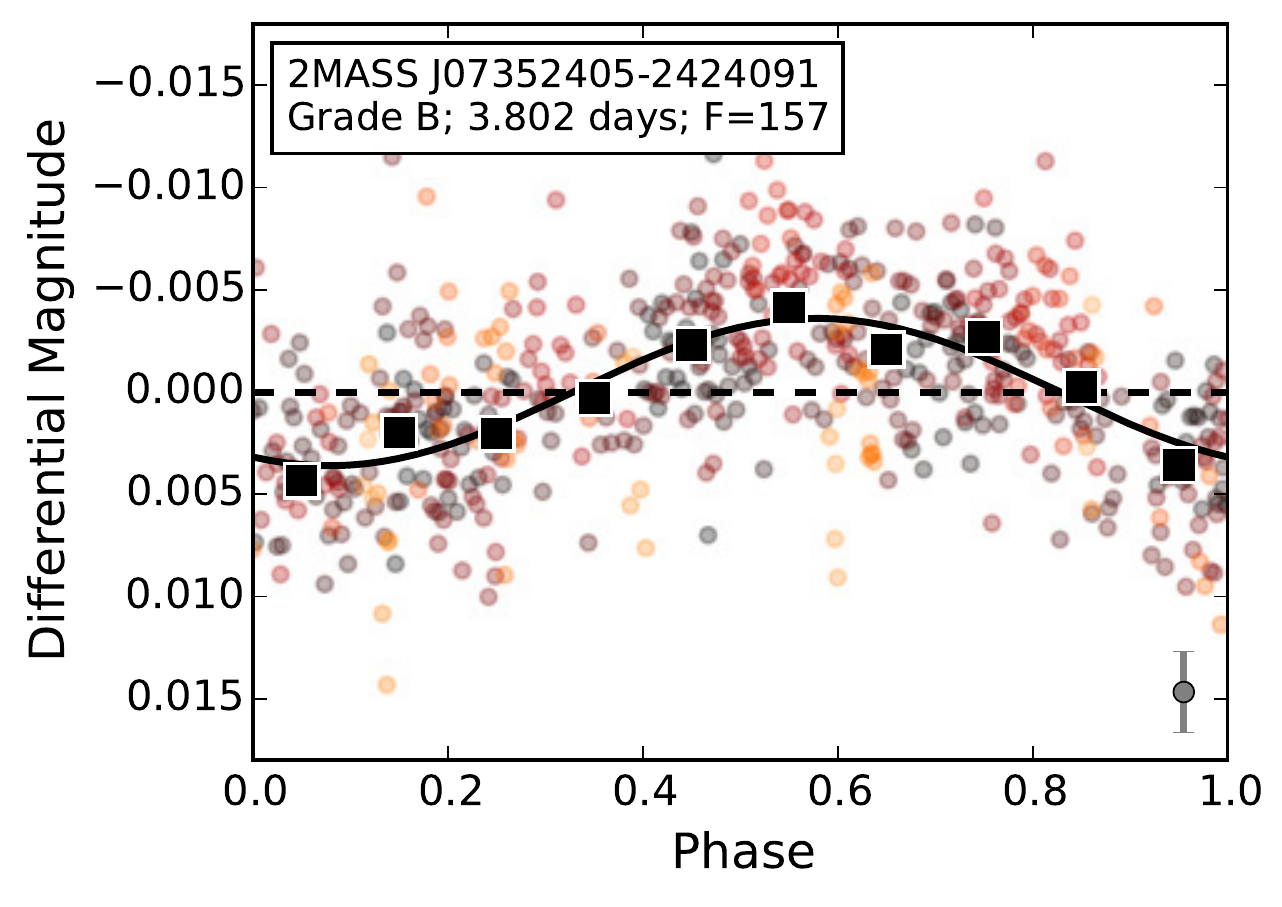}\\
\includegraphics[width=0.33\linewidth]{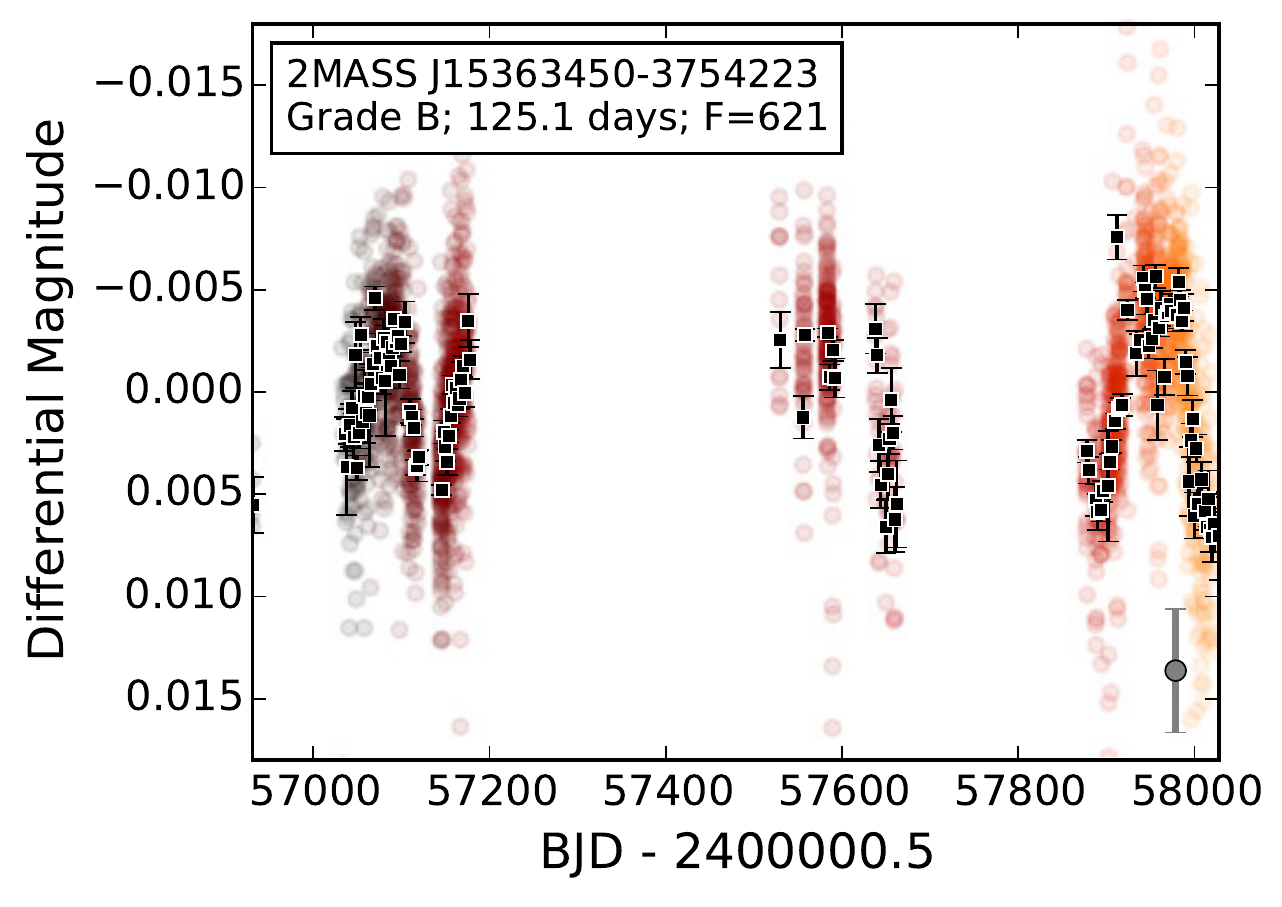}
\includegraphics[width=0.33\linewidth]{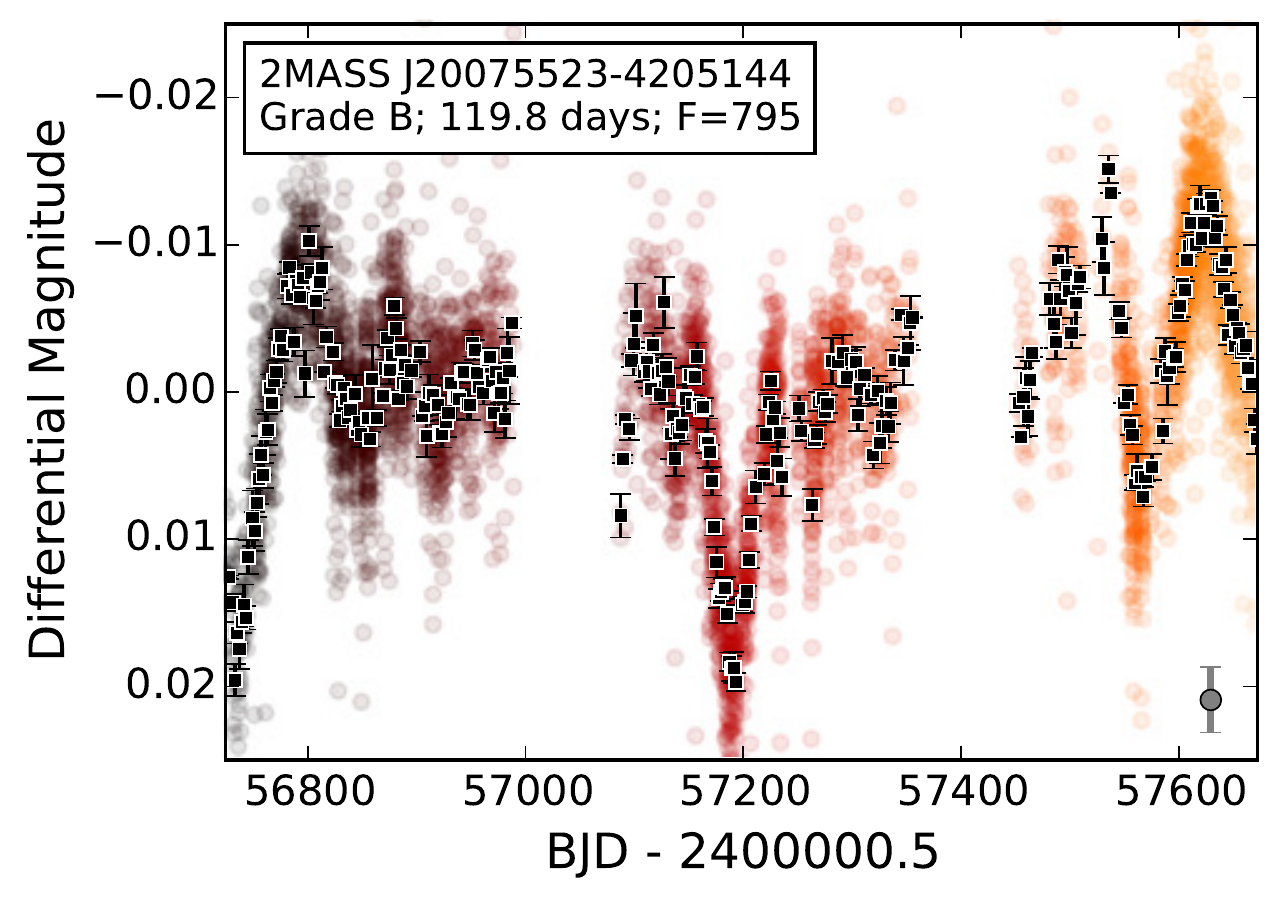}
\includegraphics[width=0.33\linewidth]{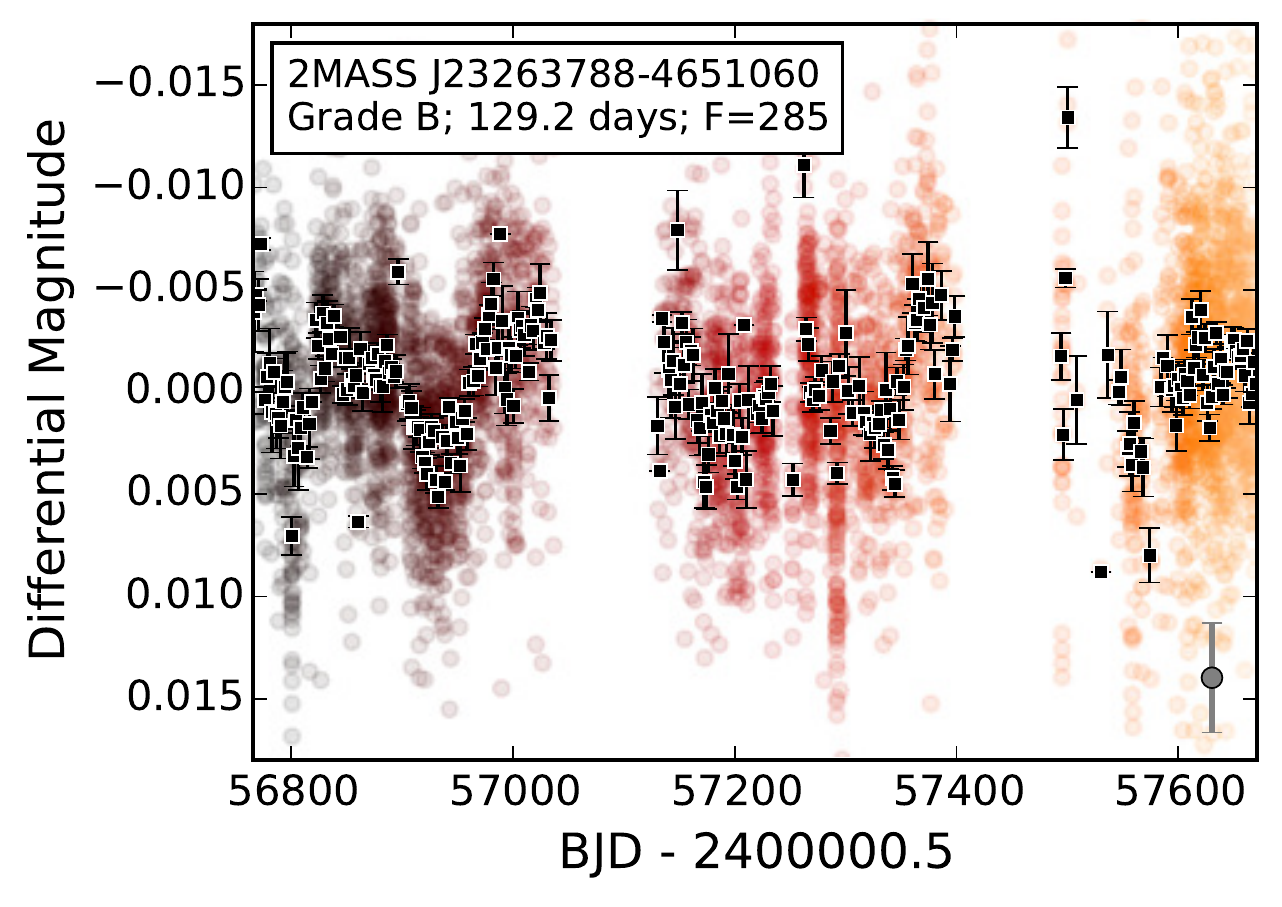}
\caption{Example lightcurves showing randomly selected grade A and B rotators, three fast and three slow rotators each. For the fast rotators, the phase-folded lightcurves are shown and for the slow rotators, time-series lightcurves are shown. Phase-folded lightcurves for all rotators and candidates, and time-series lightcurves for those with periods $\ge70$ days, are available through the IOP figure set feature. The label for each panel indicates the rating we have assigned, our adopted rotation period, and the F-test statistic (F). The color scale is based on the observation number (extending from dark red to light orange), and the median error is indicated in the bottom right corner. In the phase-folded plots, the black squares show the median magnitude in ten uniformly spaced bins; the sample scatter about the median is also plotted but is usually too small to be seen. In the time-series plots, the black squares show the median and sample scatter in two-day bins.
\label{Fig:lcs}}
\end{figure*}

 Figure \ref{Fig:massper} shows an overview of the rotation periods detected in this work. We detected $\numrotators$ rotation periods after having searched $\numall$ stars, which comprises $\numyes$ grade A rotators and $\numyesbut$ grade B rotators. We also find $\numpossibly$ candidates. Phase-folded lightcurves are shown in Figure \ref{Fig:lcs}. As in \citet{Newton2016}, we use the results from \citet{Irwin2011} to estimate that our errors in rotation period are about $10\%$. The periods and amplitudes show patterns consistent with those presented in \citet{Newton2016}, including the dearth of stars with intermediate rotation periods (approximately $10$ to $70$ days). We also find no intermediate rotators amongst the nearby M dwarfs observed by \ktwo. 
 
We note that aliases are likely present, and is some cases it is challenging to distinguish the two most likely possibilities. For example in 2MASS J06464109-2150161 and 2MASS J17281105-0143569, periods of around $1$ day and around $30$ days both provide reasonable fits to the data, and the precision and quantity of data over a single night isn't sufficient to distinguish the two. Higher cadence monitoring would be beneficial.

As in \citet{Newton2016}, sinusoidal patterns dominate the lightcurves of stars with detected rotation periods. However, aided by the lack of a monsoon season imposing summer-long gaps in, and the generally higher S/N of our southern data, we see clear departures from sinusoids in a number of cases. This could affect our best-fitting rotation periods, which we suggest on the basis of our $89$ day period for Proxima, in contrast to the $83$ period consistently reported in the literature \citep{Benedict1998, Wargelin2017}. Using Gaussian processes to model the stellar variability \citep[e.g.][]{Cloutier2016, Angus2018} for cases of unique interest would likely be beneficial.

\begin{figure*}
\centering
\includegraphics[width=0.85\linewidth]{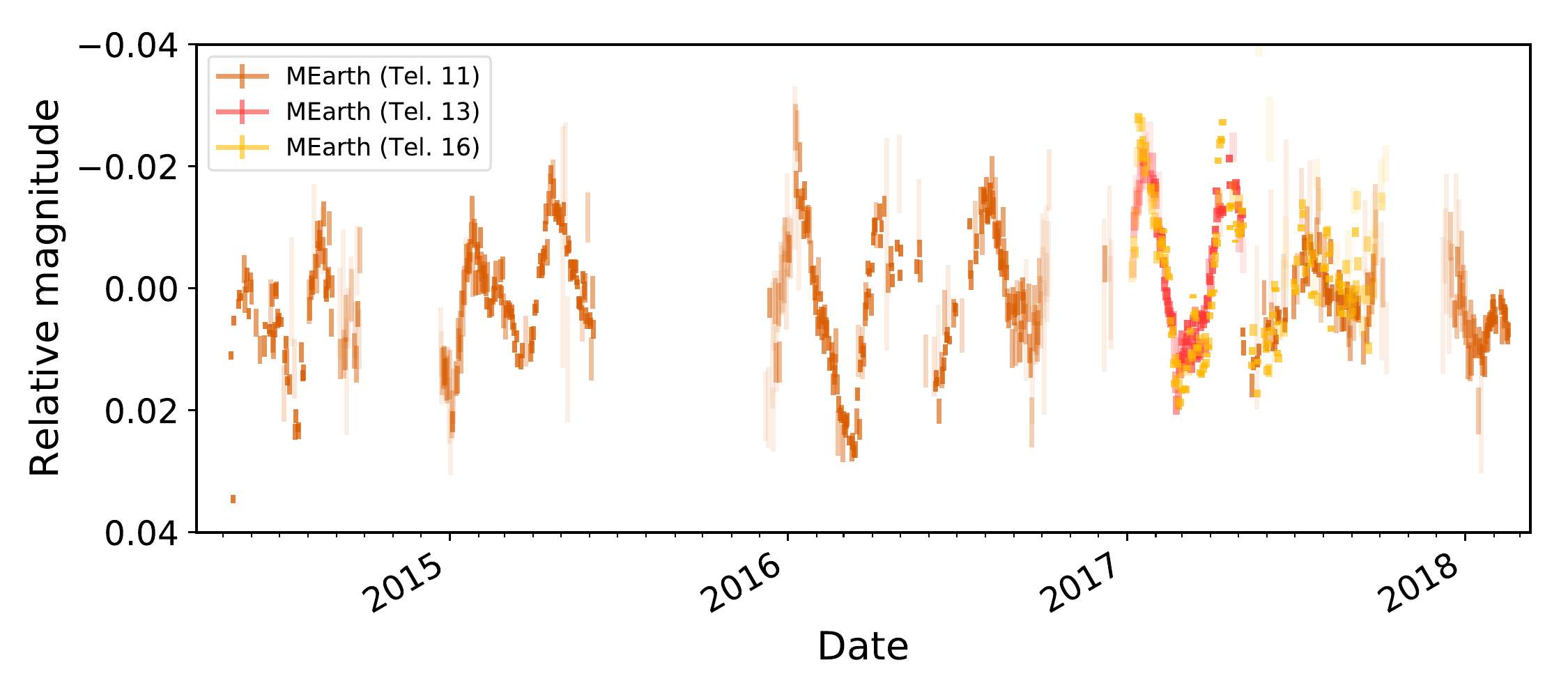}
\caption{Lightcurve of Proxima Centauri from MEarth-South. The MEarth dataset has been median combined into one day bins, with the error bars representing the standard error on the mean (calculated as 1.48 times the median absolute deviation). Data from three telescopes are shown (orange, pink, and yellow). The magnitude offsets are derived by our fitting code.
\label{Fig:prox}}
\end{figure*}

Amongst the stars in our sample is the 
mid M dwarf Proxima Centauri, the nearest star to us. \citet{Benedict1998} reported an $83$ day rotation period for Proxima using the \emph{Hubble Space Telescope} Fine Guidance Sensors, which was confirmed by \citet[][82.5 days]{Kiraga2007} and others. Recently, \cite{Wargelin2017} identified a $7$ year activity cycle for Proxima based on a $15$ years of photometry from ASAS \citep{Pojmanski2002}. Our lightcurve of Proxima is shown in Fig. \ref{Fig:prox}. We measure a rotation period of $89$ days.

\begin{figure*}
\centering
\includegraphics[width=0.85\linewidth]{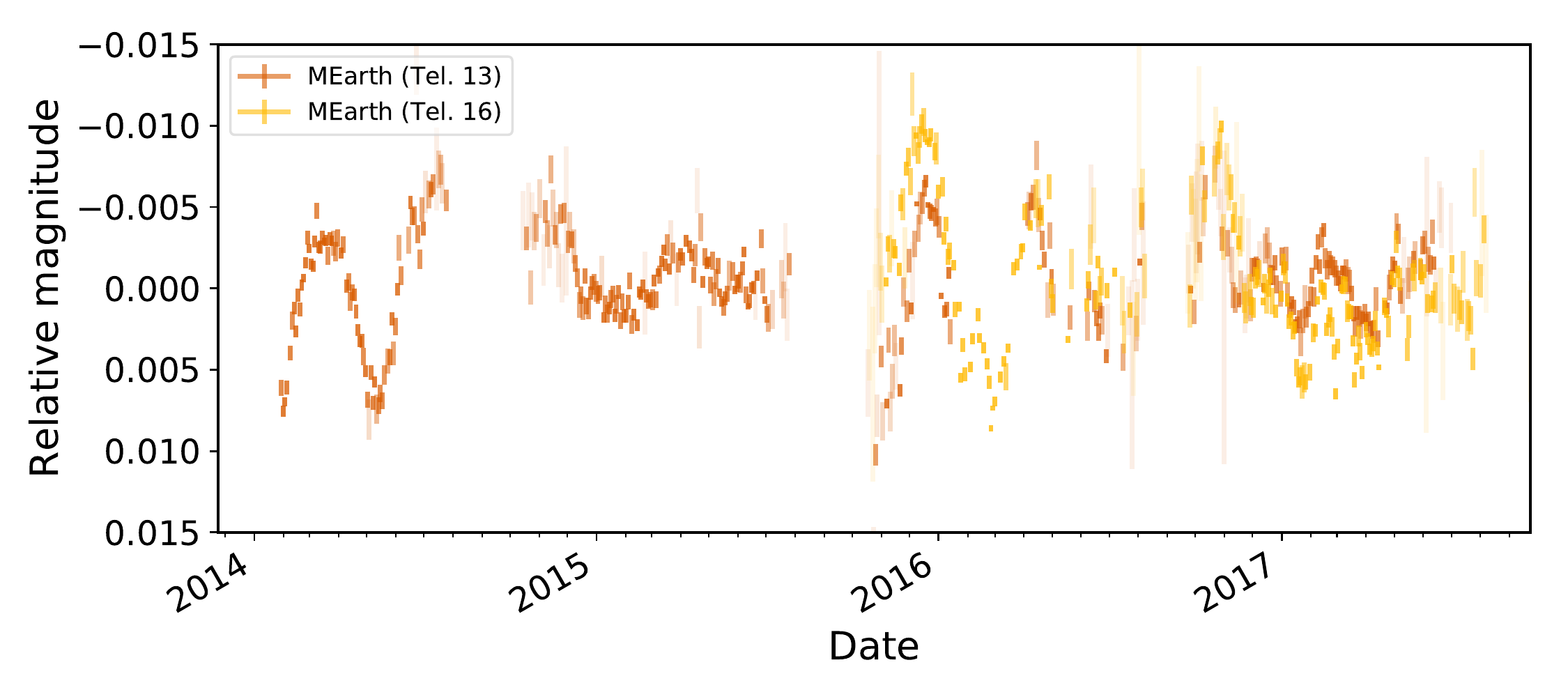}
\caption{Lightcurve of GJ 1132 from MEarth-South. The MEarth dataset has been median combined into one day bins, with the error bars representing the standard error on the mean (calculated as 1.48 times the median absolute deviation). Data from two telescopes are shown (orange and yellow). The magnitude offsets are derived by our fitting code. Substantial spot evolution is seen on timescales similar to the rotation period. 
\label{Fig:gj1132}}
\end{figure*}

Our sample also includes the planet hosts GJ 1132 \citep{Berta-Thompson2015} and LHS 1140 \citep{Dittmann2017}. Rotation periods for both of these stars, measured from MEarth-South data, were presented in those works. They are: $125$ days for GJ 1132 and $131$ days for LHS 1140. For GJ 1132, \citet{Cloutier2016} measure $122^{+6}_{-5}$ days using Gaussian process modeling applied to the MEarth data then available (the first half of the current dataset, through BJD 2457285). The period we calculate based on the current dataset is $130$ days for GJ 1132, within the approximately $10\%$ error bars. LHS 1140's period is unchanged. The lightcurve of GJ 1132 is shown in Figure \ref{Fig:gj1132}, and demonstrates substantial spot evolution on timescales similar to the rotation period.

\subsection{Period recovery rates}

\begin{figure}
\begin{centering}
\includegraphics[width=\linewidth]{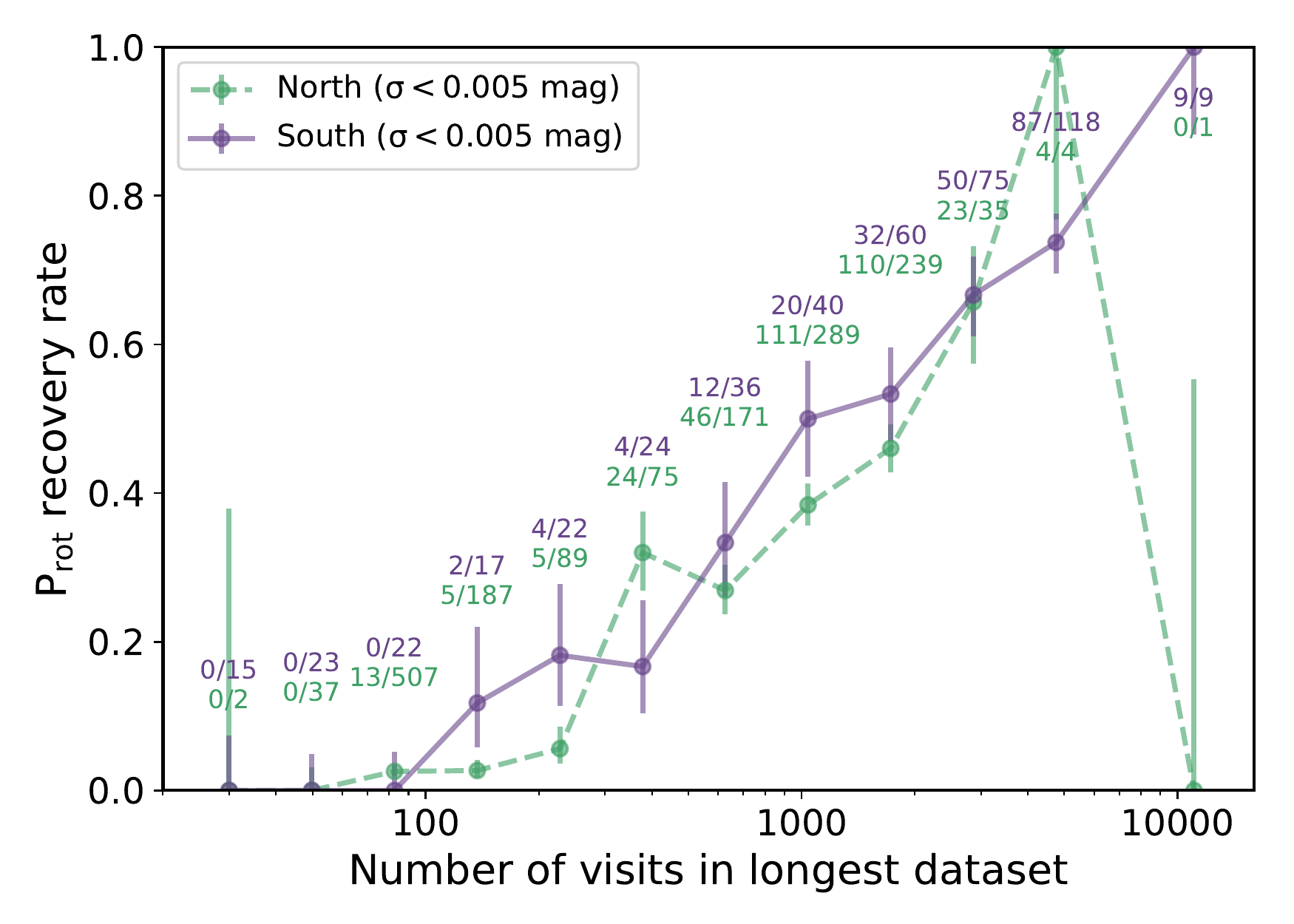}
\includegraphics[width=\linewidth]{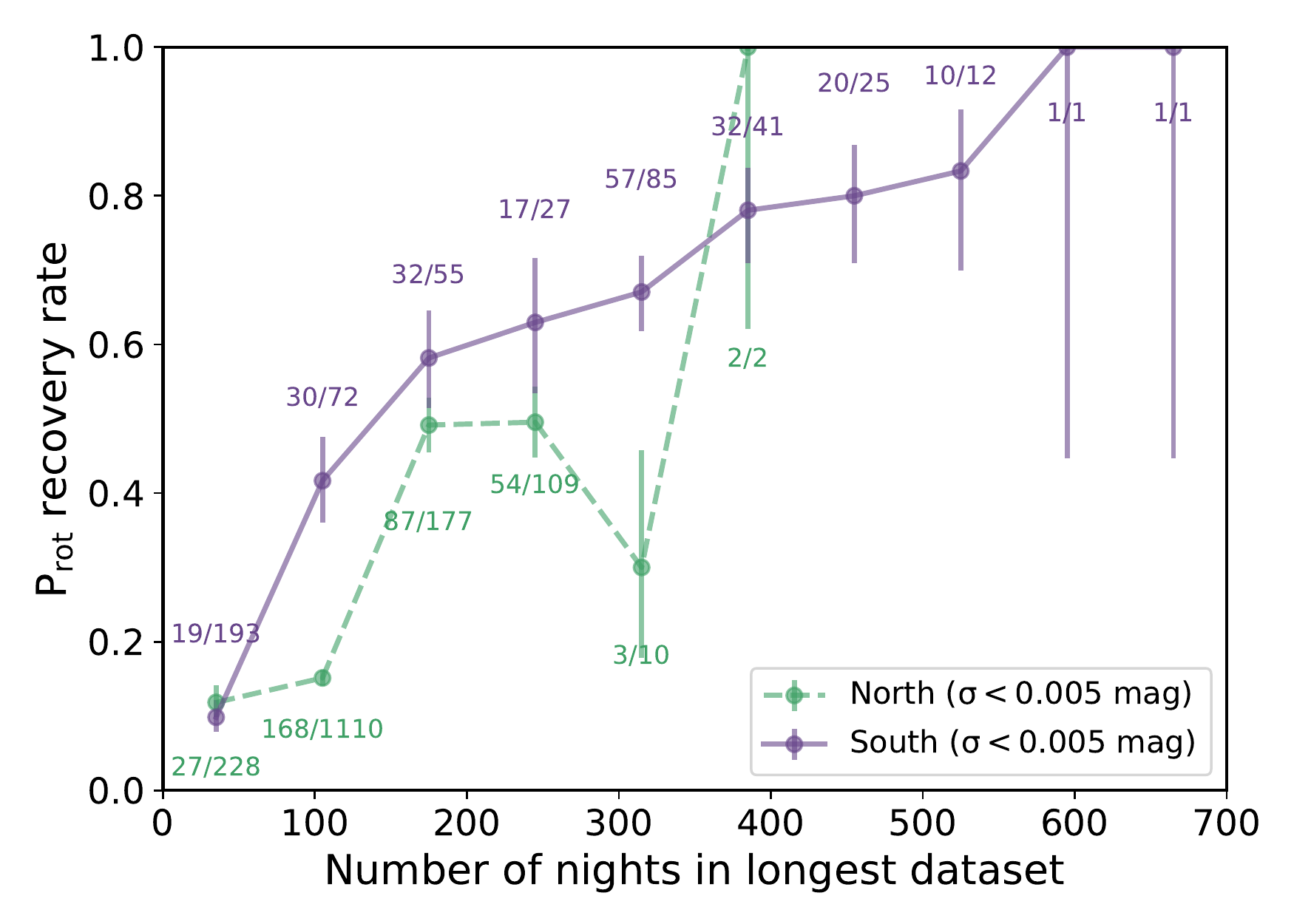}
\end{centering}
\caption{Recovery rates for stars with different length data sets. The recovery rate is the fraction of surveyed stars for which a period was detected, either Grade A or Grade B.  We show the southern sample (purple) and the northern sample from \citet[][teal]{Newton2016}. For the northern sample, the number of observations as of the date limit used in \citet{Newton2016} are shown. The numbers of stars in each bin are indicated on the plot, with the number for the south above the number for the north. The top panel shows recovery rate as a function of the total number of visits in the longest season of observations. The bottom panel shows the recovery rate as a function of the total number of separate days included in the longest season. GJ 1132 is not included in the top panel, as it has twice the number of visits of any other star in our sample. 
\label{Fig:recovery} }
\end{figure}

We separate the data into evenly spaced bins based on length of dataset and calculate the recovery rate in each bin, shown in Fig. \ref{Fig:recovery}. If stars were observed with both MEarth-North and MEarth-South, they are categorized as being part of the southern sample. We consider only stars with radii between $0.15$ and $0.33$ $\rsun$. The upper limit is set by the fiducial radius limit for MEarth; the lower limit is a result of our decreasing sensitivity to rotation periods at lower masses due to precipitable water vapor variations (discussed in the following section). We assume that period detections represent independent binomial trials, and use the \texttt{statsmodel} package \citep{Seabold2010} to calculate confidence limits for a binomial proportion. The error bars in Fig. \ref{Fig:recovery} represent the $68\%$ confidence interval. 

We find a striking positive correlation between the length of the dataset and the fraction of stars in which periods are detected, with the recovery rate reaching $80\%$ for datasets with observations on $\geq350$ separate nights (bottom panel).
 
In \citet{Newton2016}, our overall recovery rate in the subset of northern stars termed the ``statistical sample'' was $47\pm3\%$\footnote{The error bar here, calculated as the $68\%$ confidence interval on the binomial proportion, and differs from that in \citet{Newton2016}.}. The statistical sample was defined as all stars with $>1200$ visits in a single lightcurve and median error per visit $< 0.005$ mag. Considering the same constraints in the South, our recovery rate is $66\pm3\%$. 

In the following section, we will consider stars for which the median error in one visit is $<0.005$ mag and for which observations were acquired on $\geq350$ individual nights 
(having $350$ days of observations generally involves an observing baseline of around $1000$ days, given weather and seasonal observing gaps). 
We found that this better represented our by-eye assessments of whether a rotation period had been detected than the $>1200$ visit requirement.
We note that we regularly drop stars from the survey when they reach this limit; this choice was set by sensitivity to planets, but we coincidentally find that $350$ days is a reasonable cut-off for rotation period monitoring. In this subset of stars, our recovery rate is $79\pm5\%$.

\subsection{Stars without detected periods}\label{Sec:nondetections}

\begin{figure*}
\centering
\includegraphics[width=0.33\linewidth]{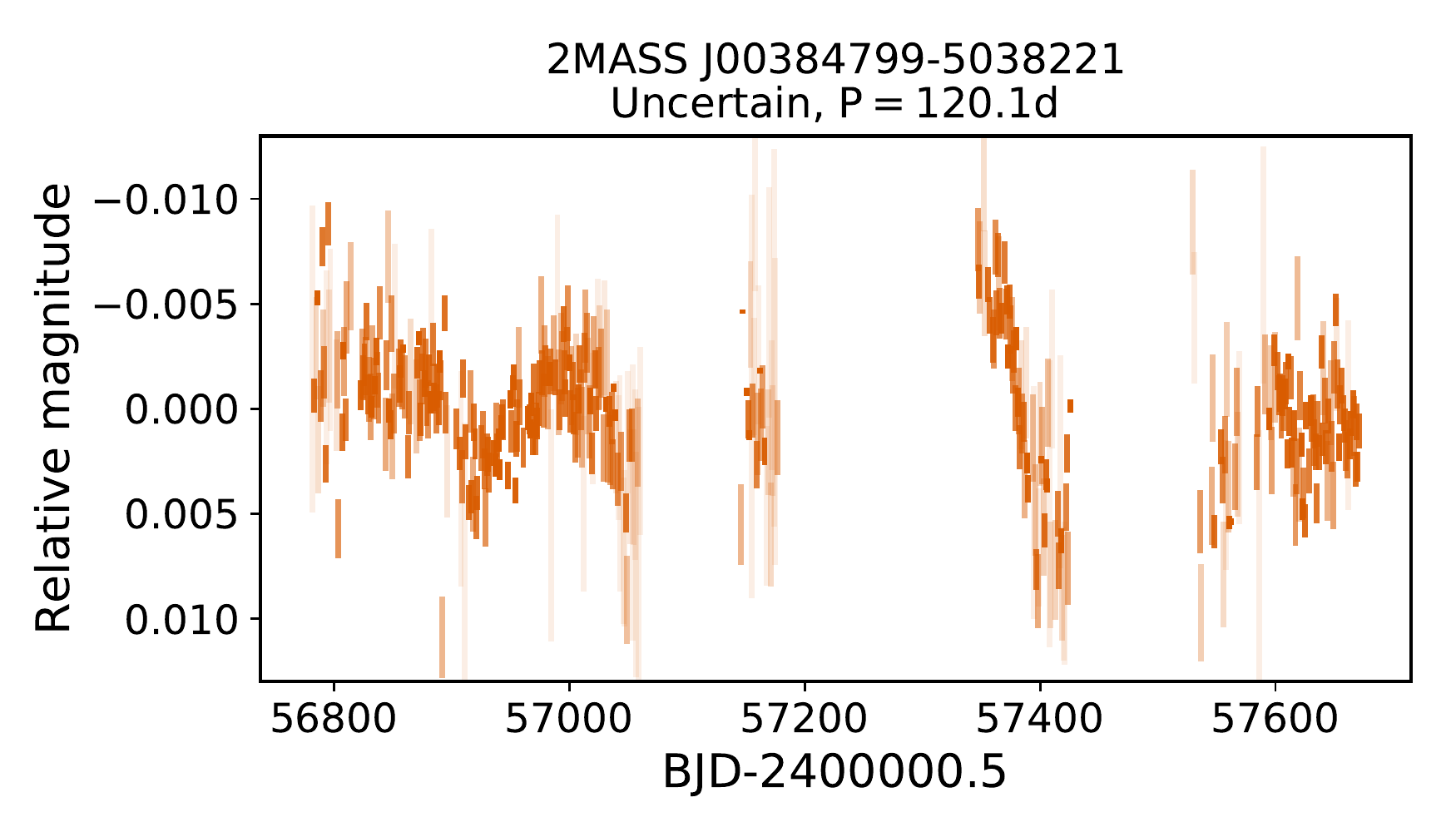}
\includegraphics[width=0.33\linewidth]{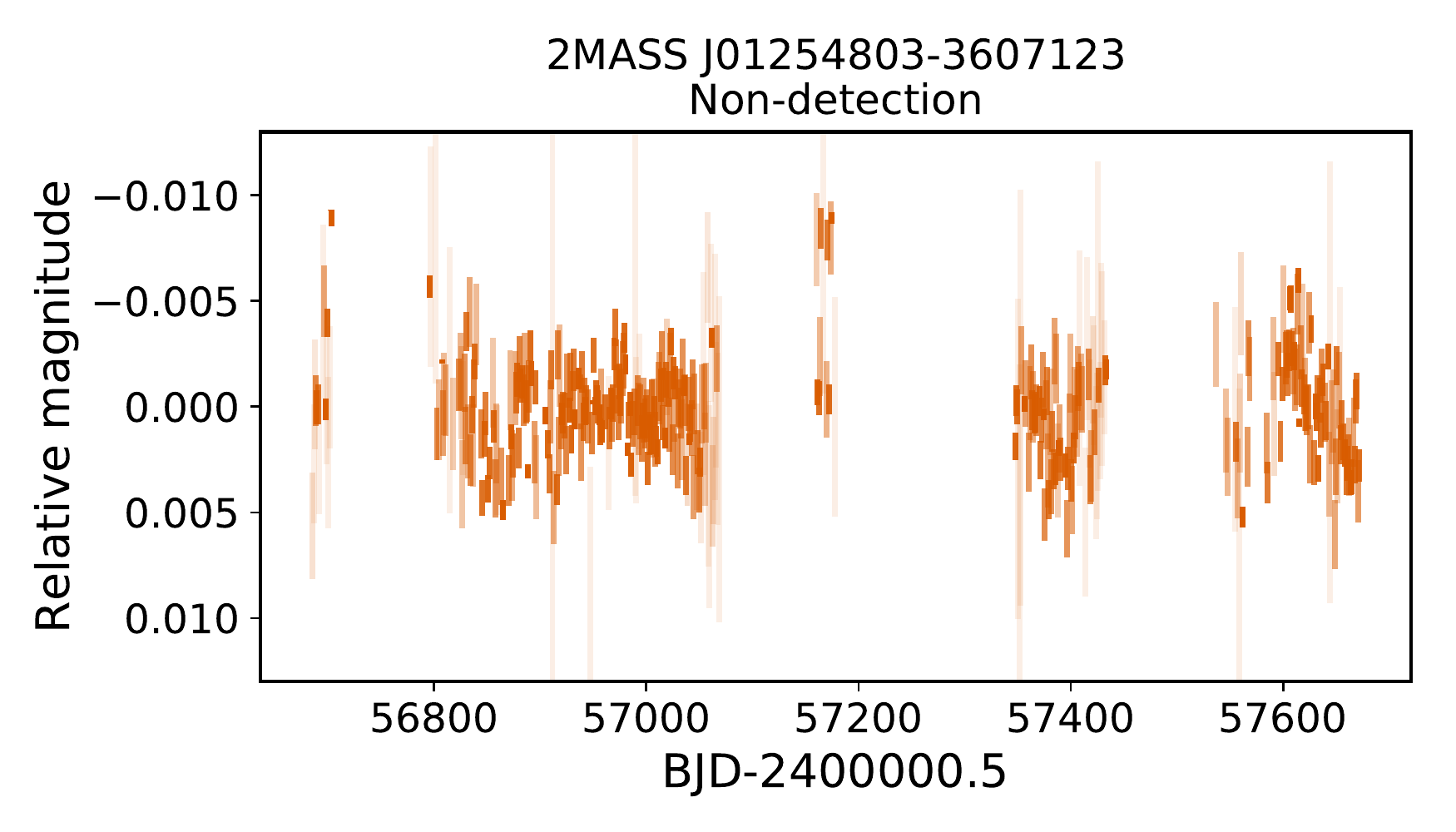}
\includegraphics[width=0.33\linewidth]{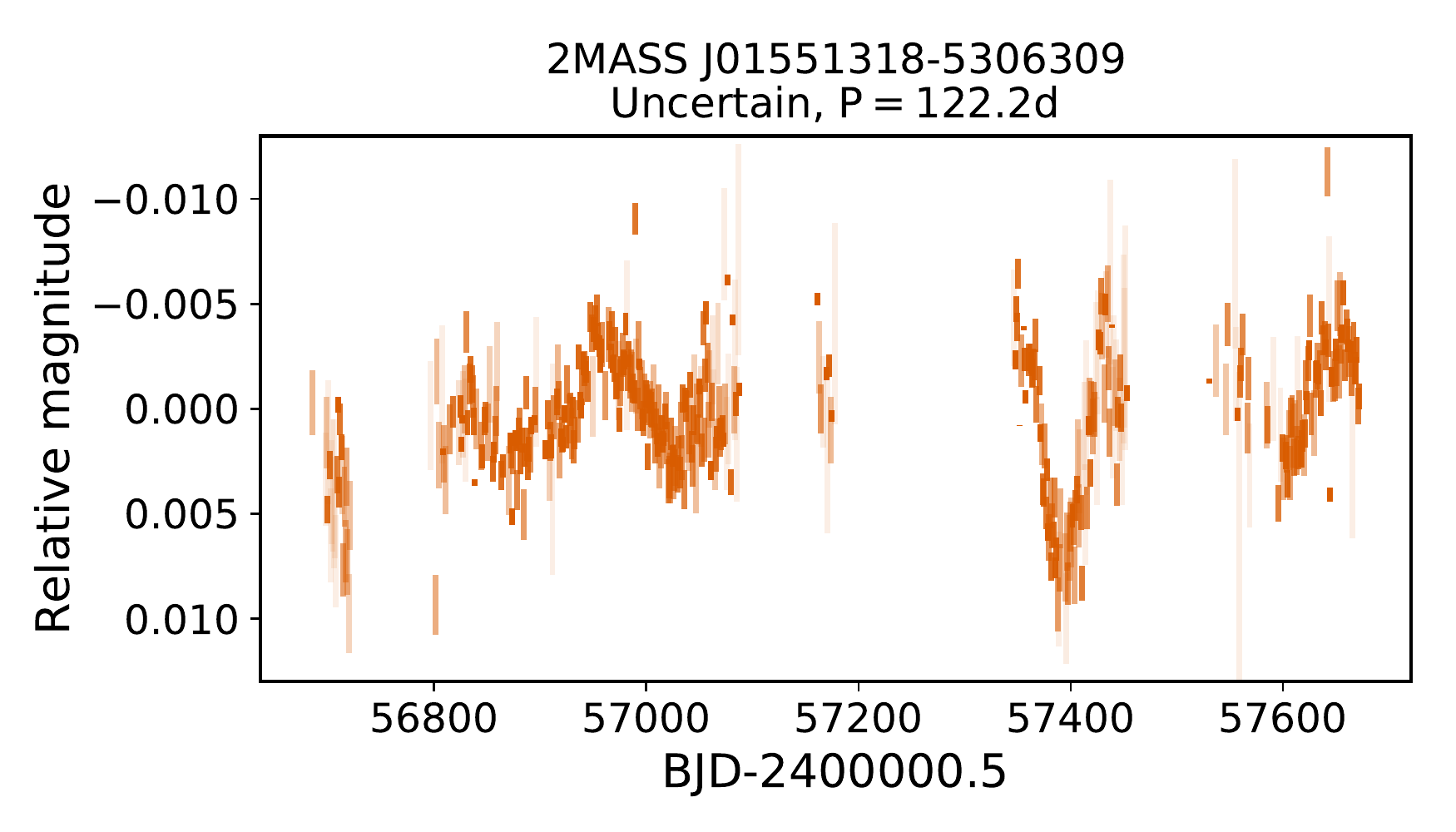}\\
\includegraphics[width=0.33\linewidth]{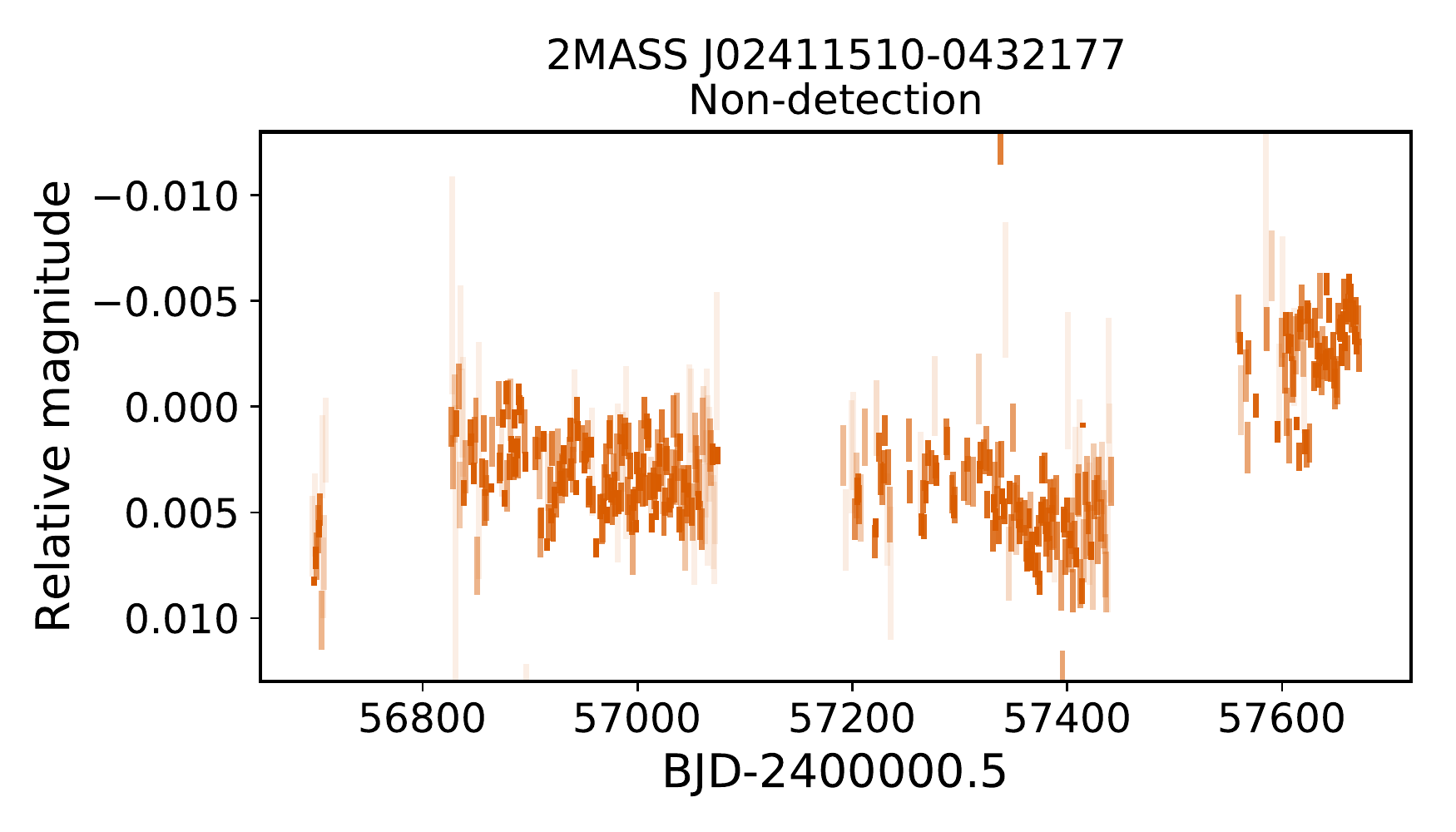}
\includegraphics[width=0.33\linewidth]{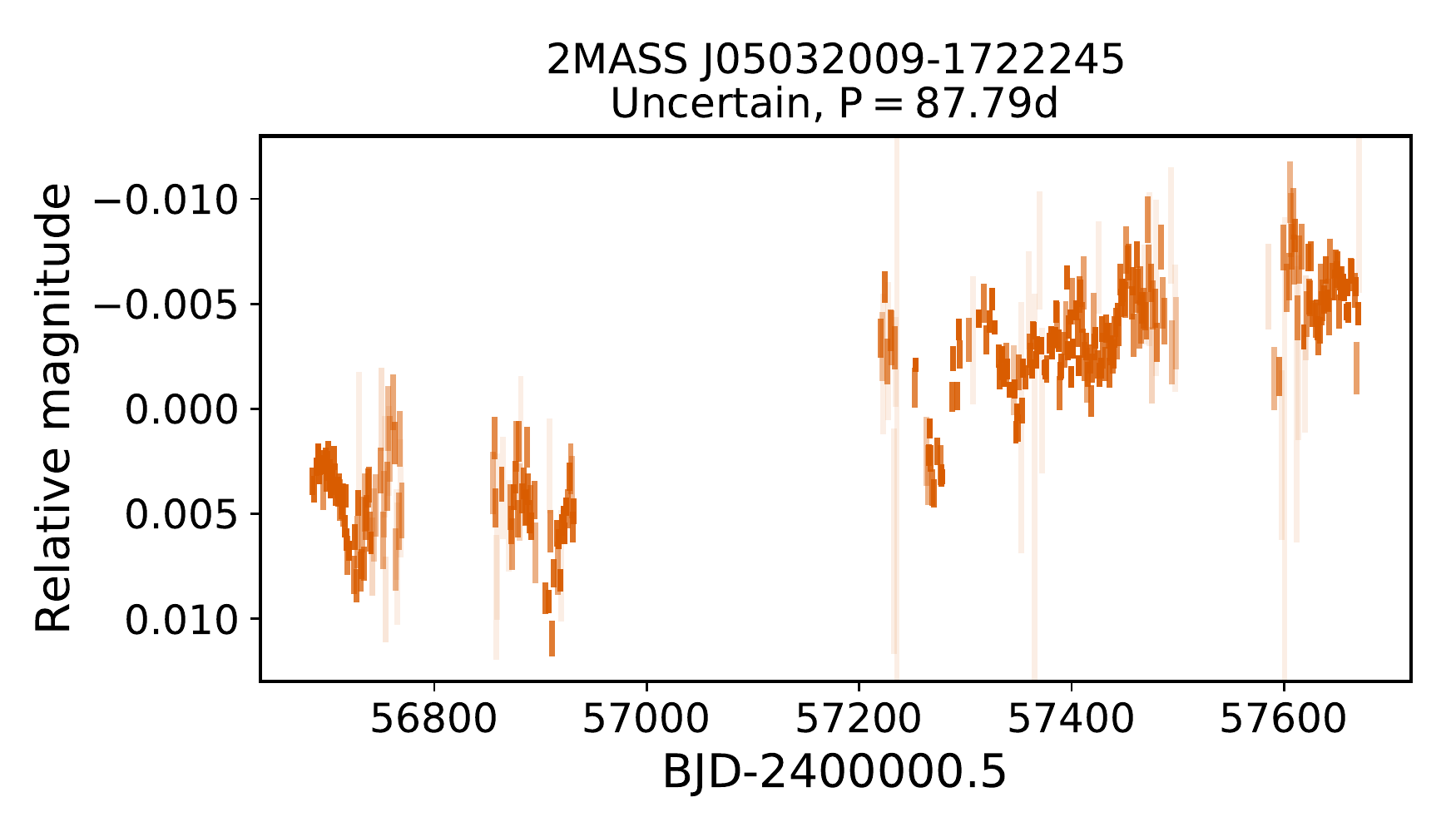}
\includegraphics[width=0.33\linewidth]{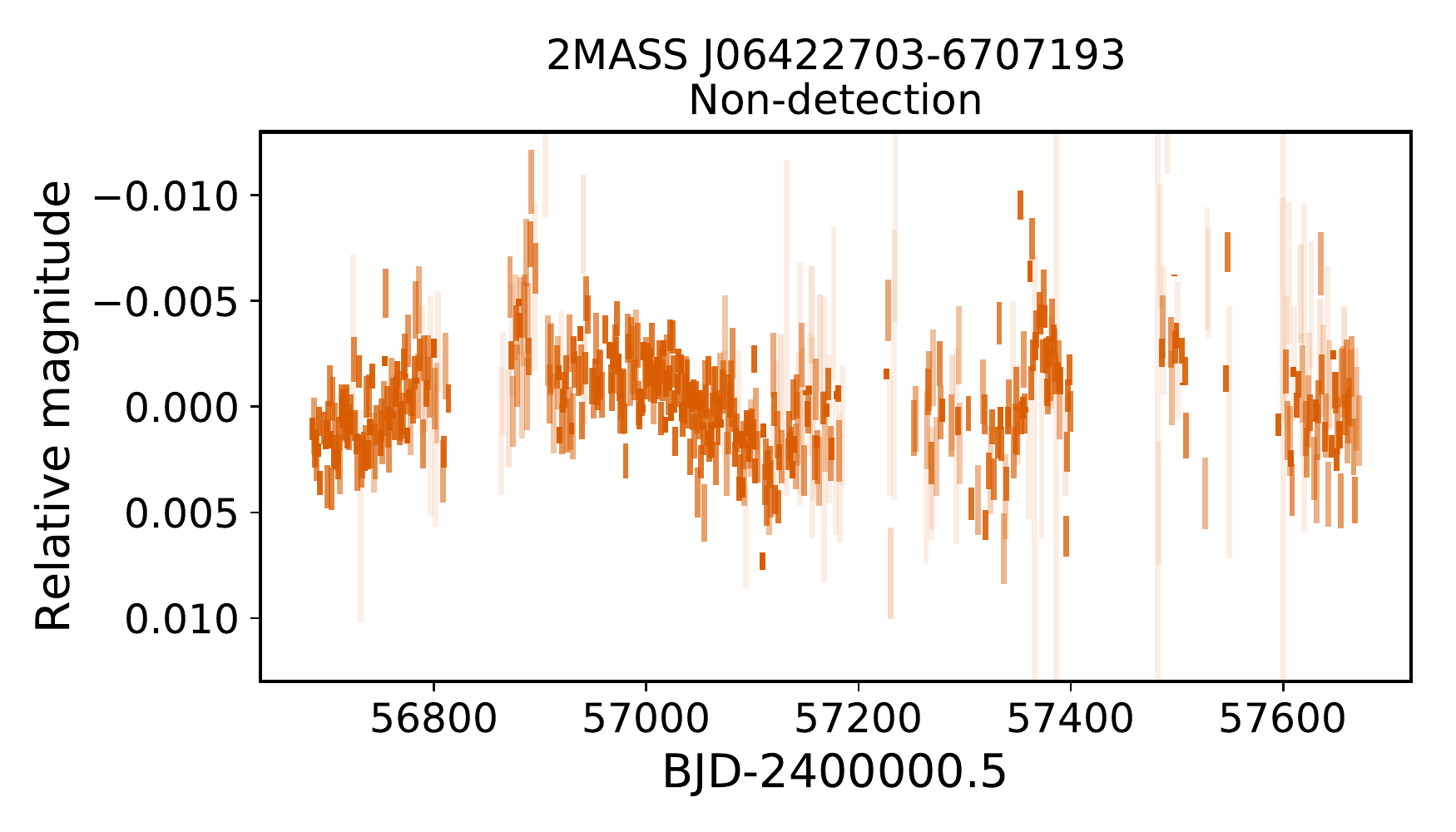}\\
\includegraphics[width=0.33\linewidth]{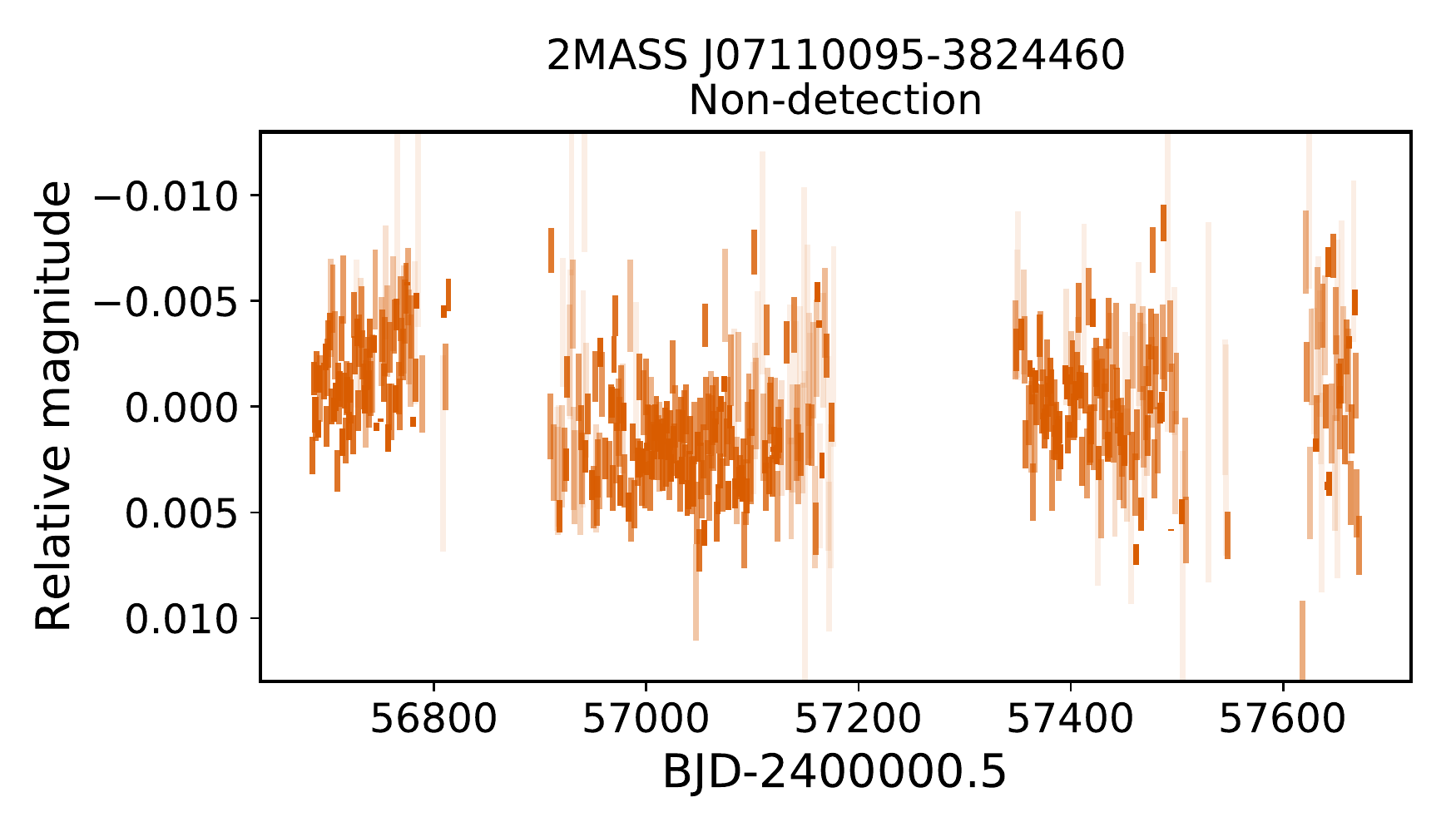}
\includegraphics[width=0.33\linewidth]{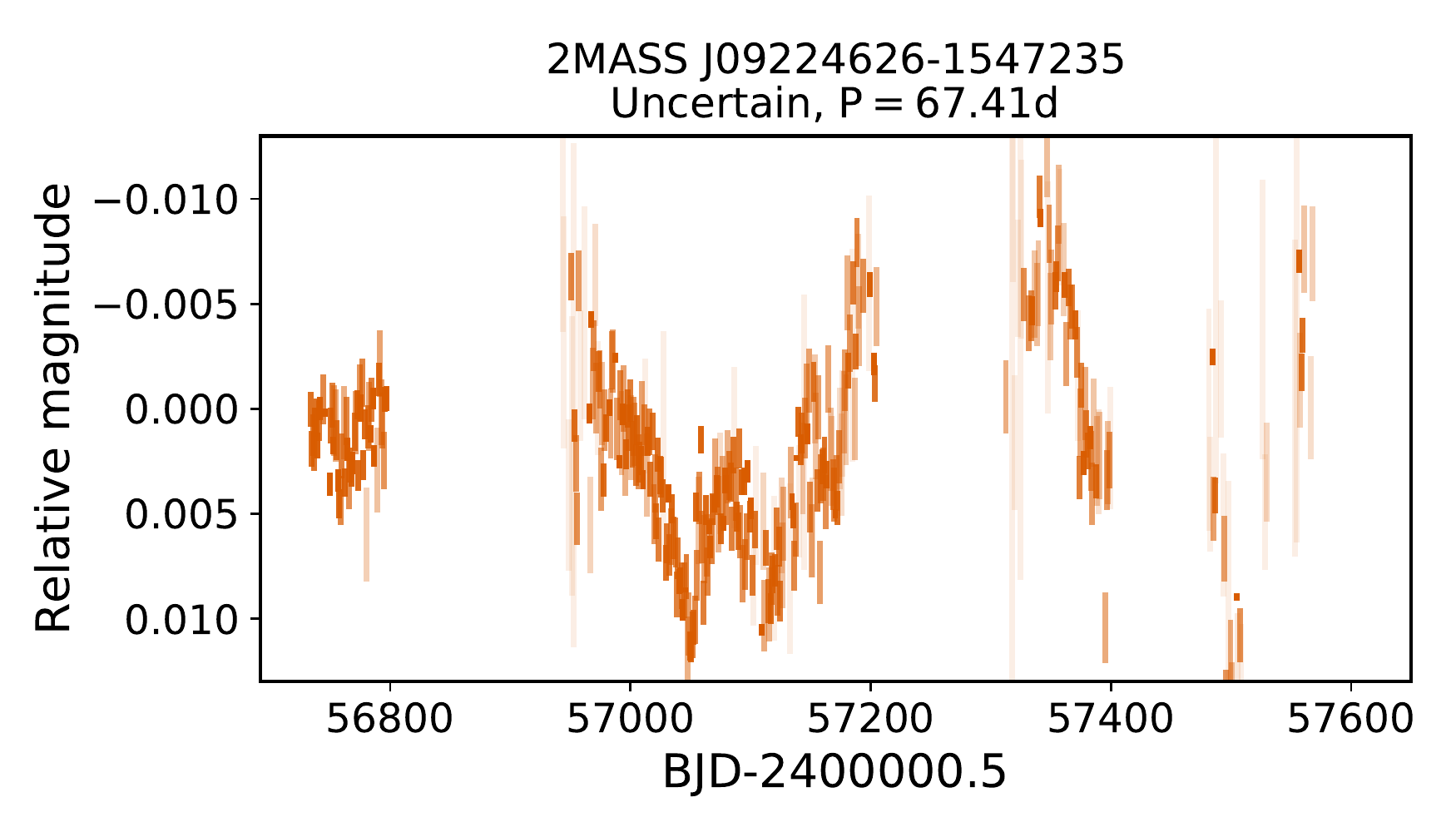}
\includegraphics[width=0.33\linewidth]{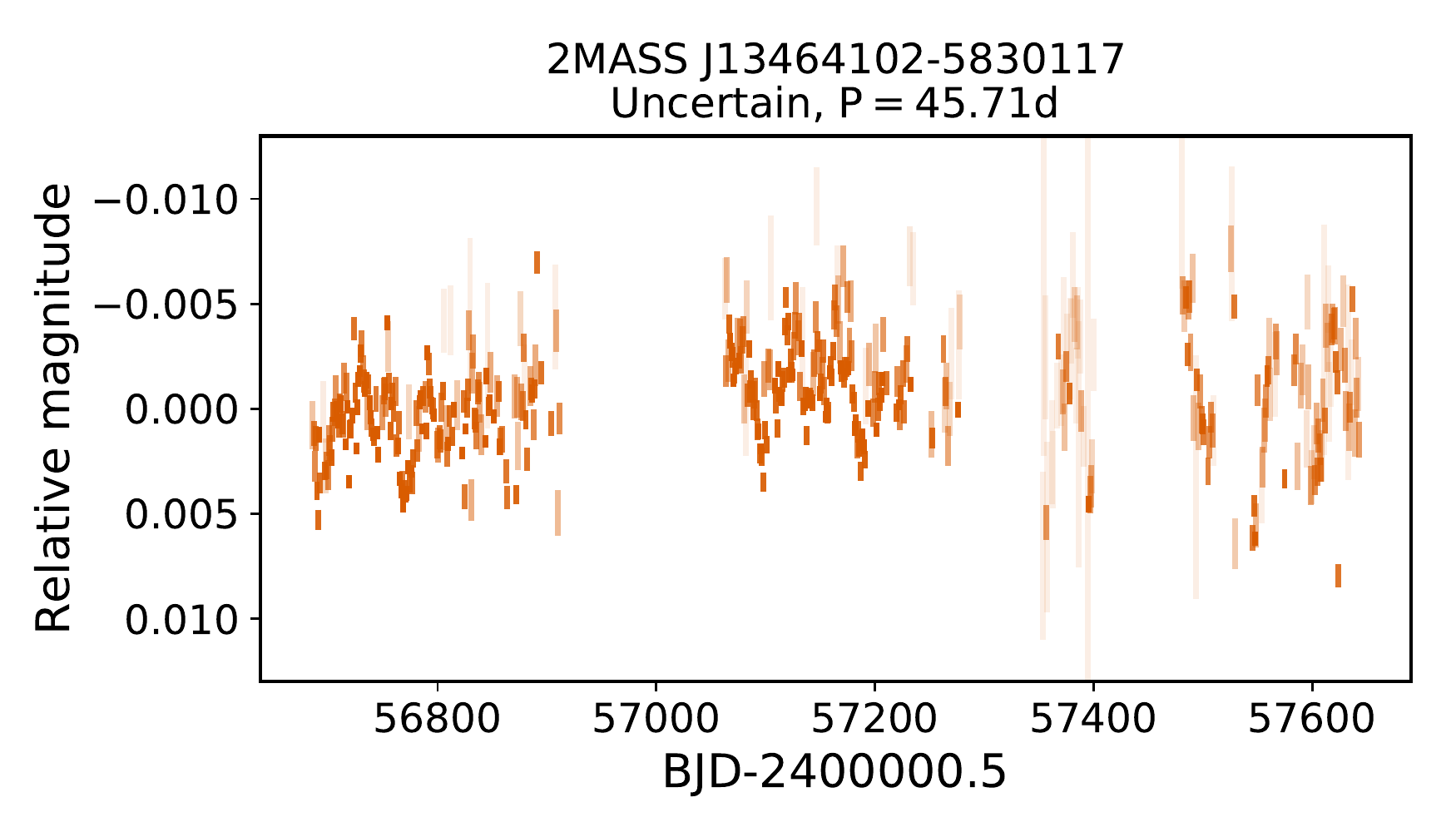}\\
\includegraphics[width=0.33\linewidth]{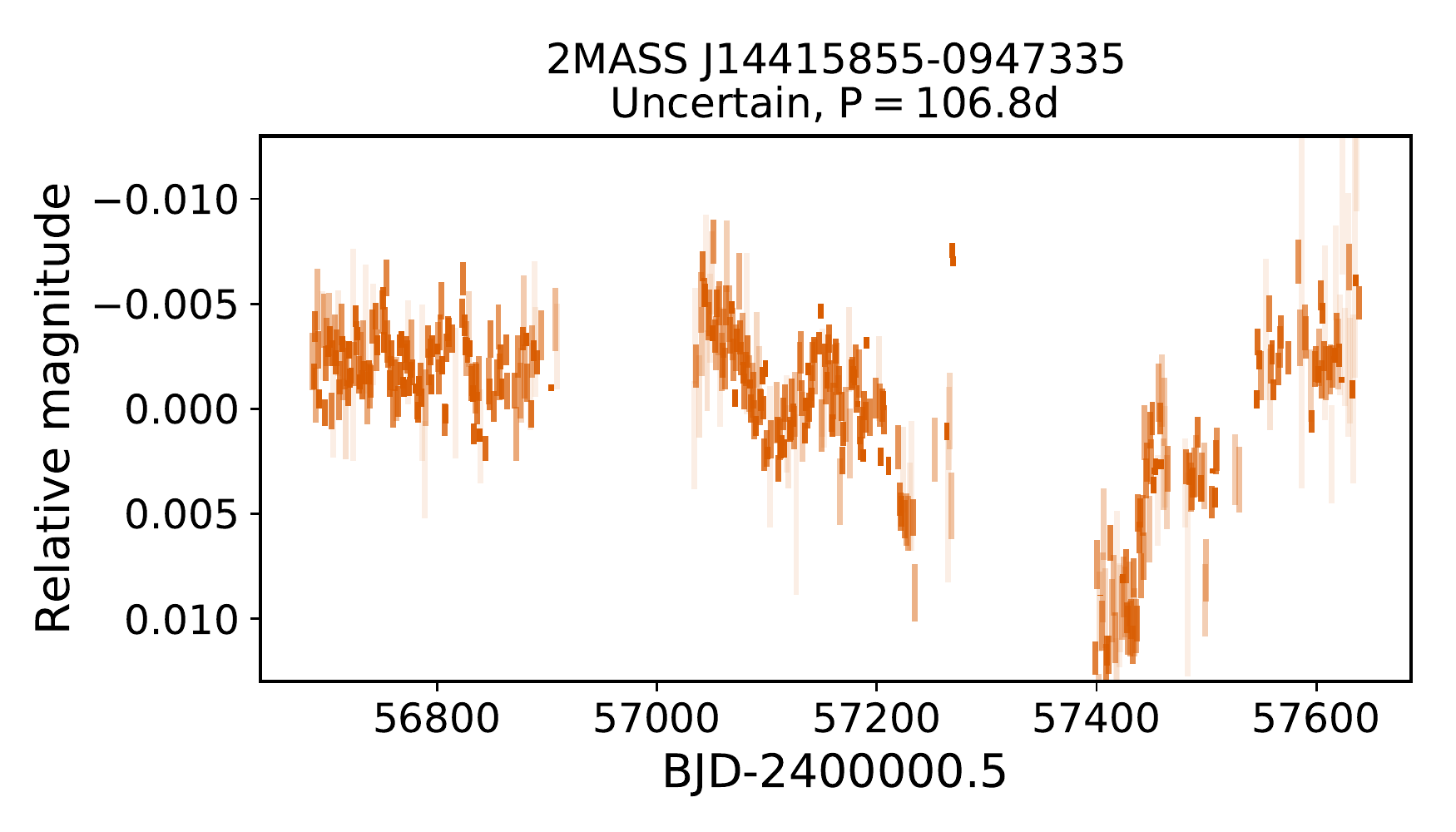}
\includegraphics[width=0.33\linewidth]{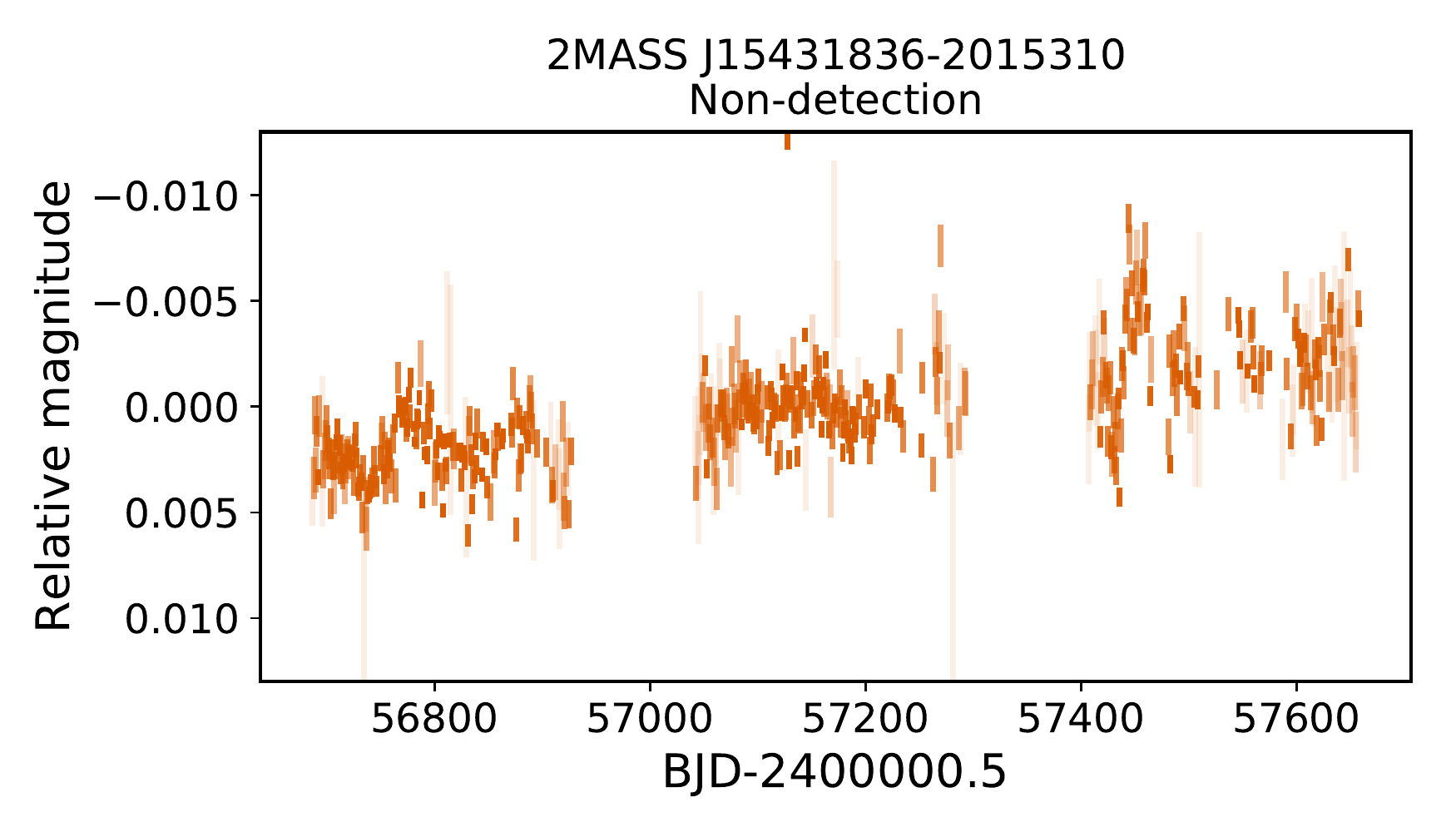}
\includegraphics[width=0.33\linewidth]{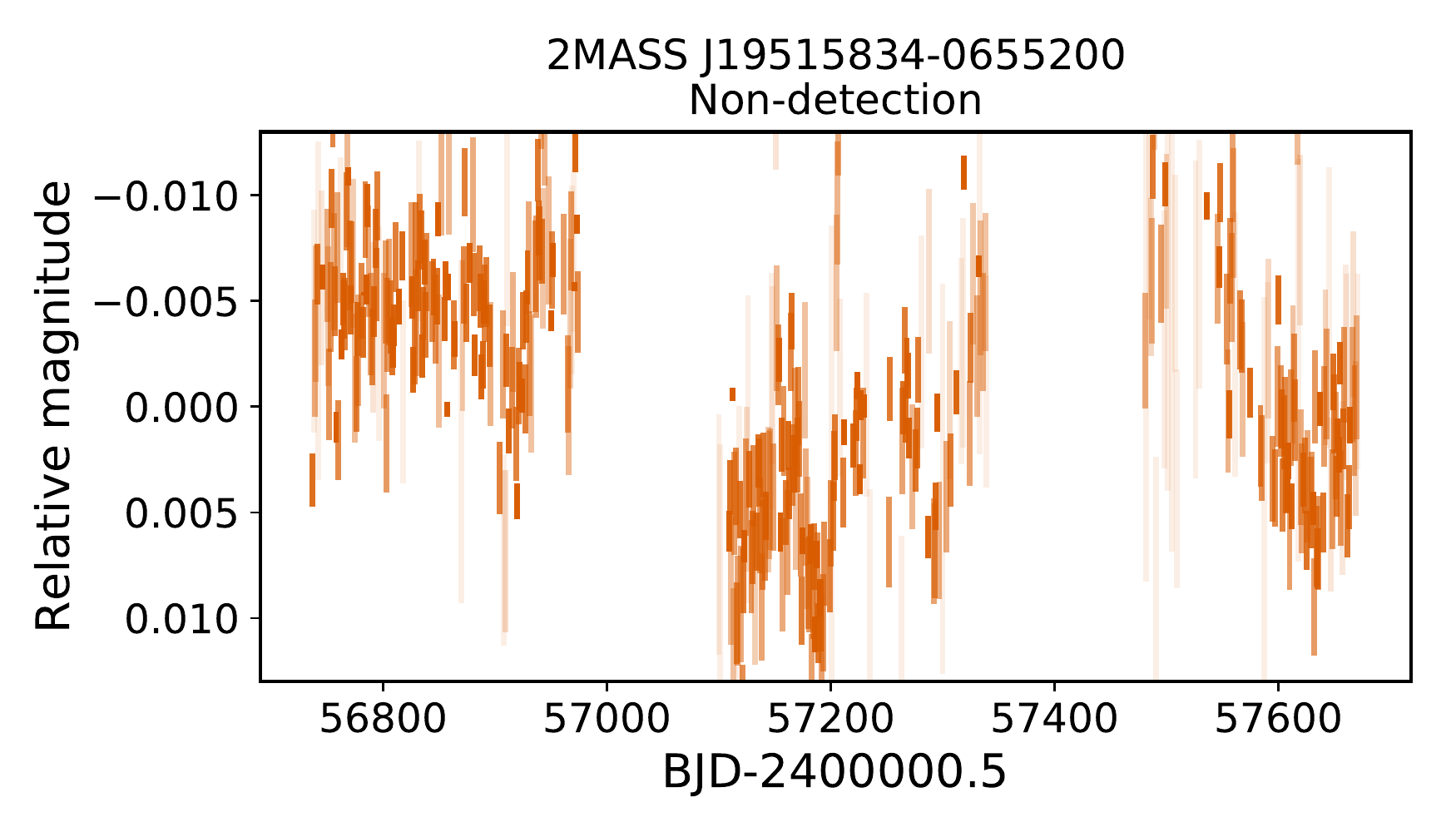}\\
\includegraphics[width=0.33\linewidth]{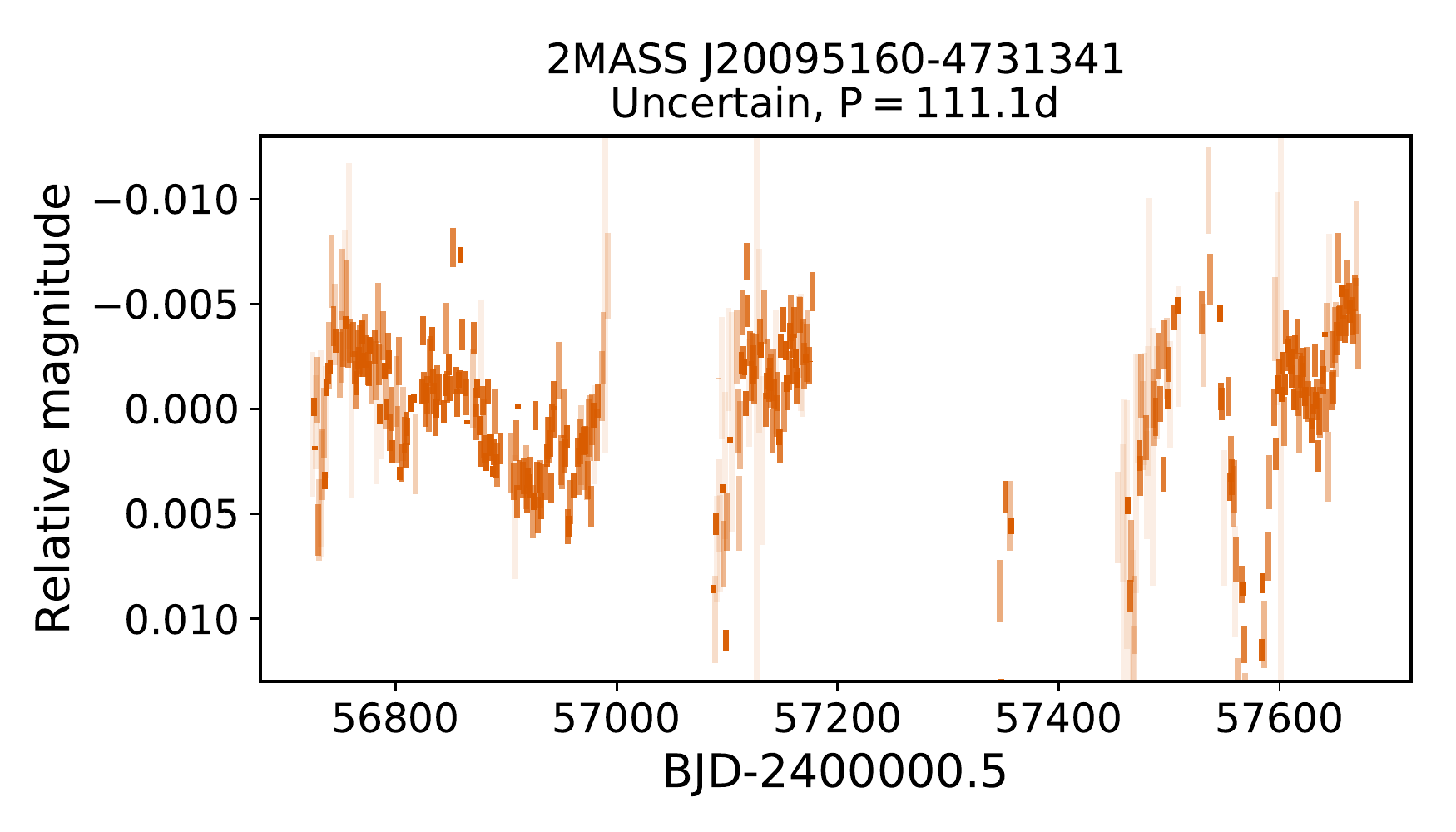}
\includegraphics[width=0.33\linewidth]{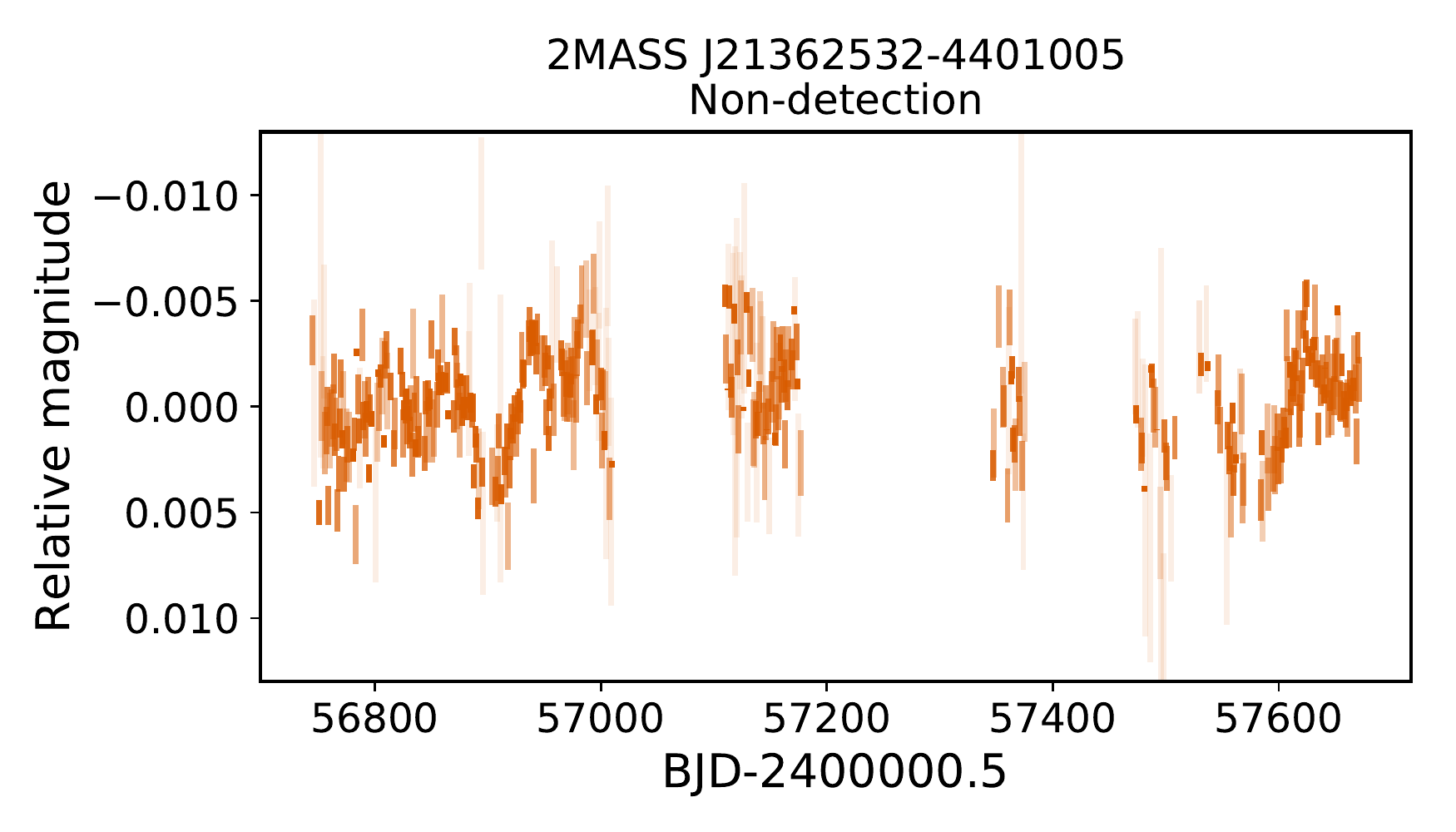}
\includegraphics[width=0.33\linewidth]{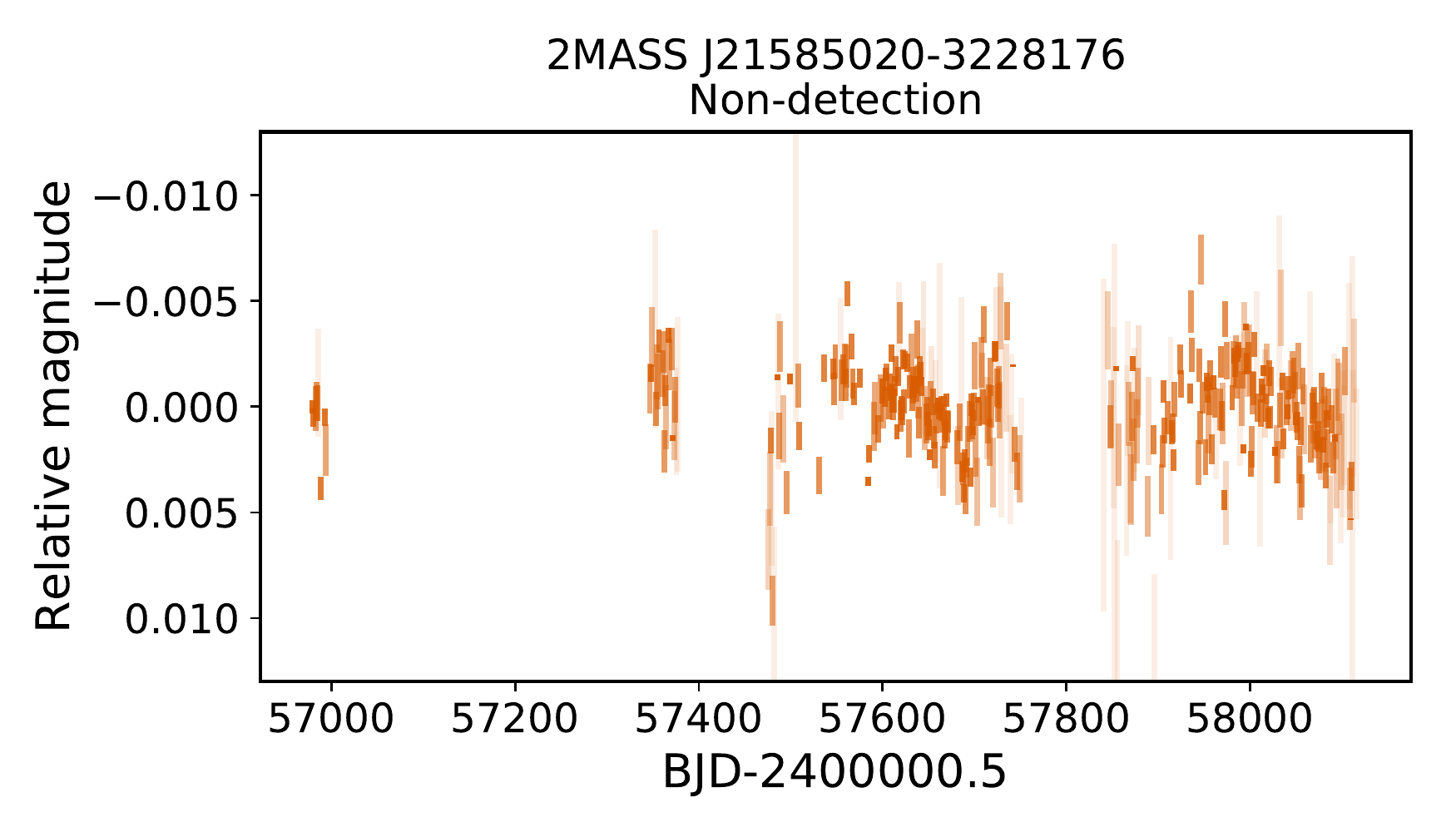}\\
\includegraphics[width=0.33\linewidth]{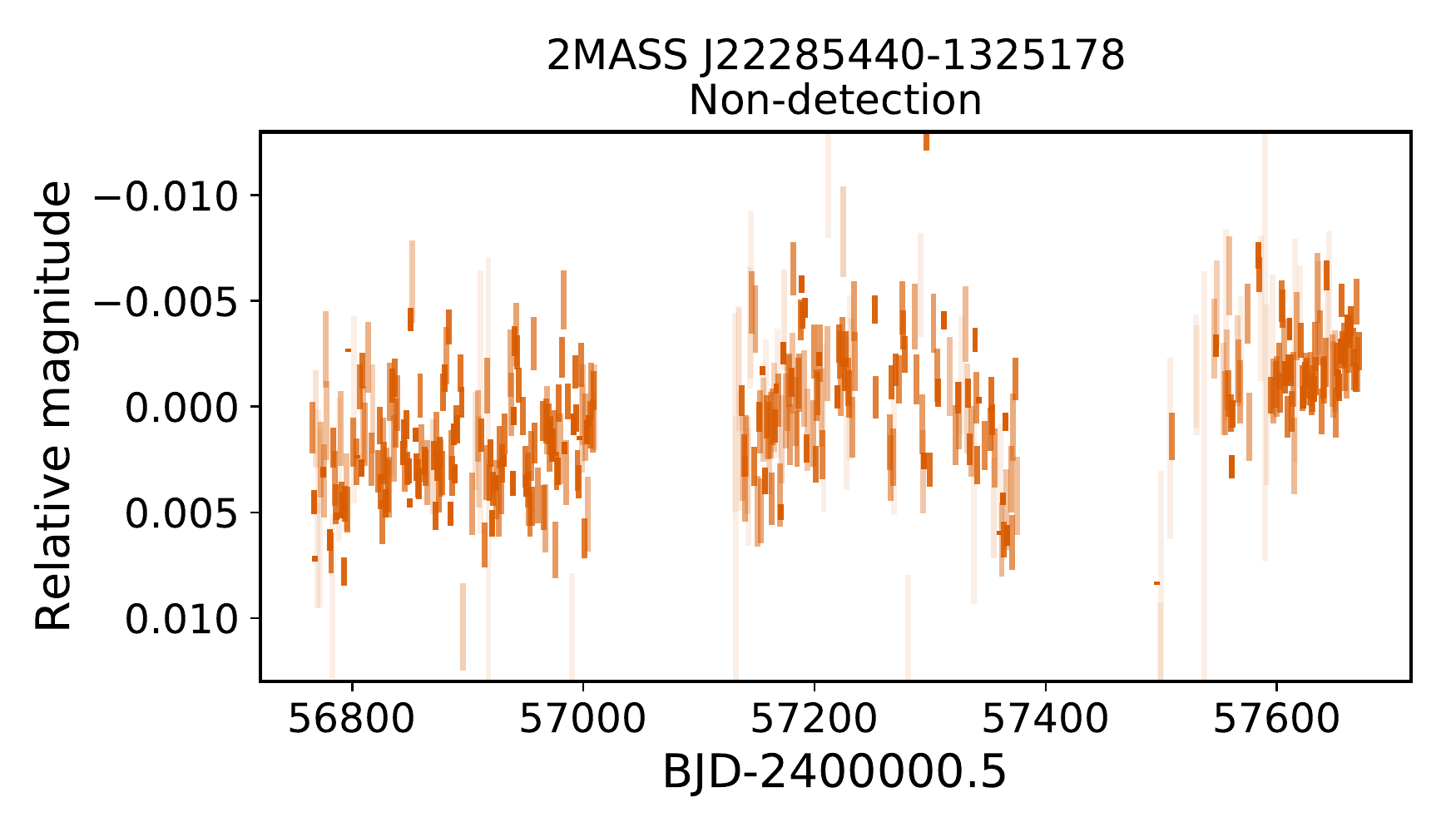}
\caption{Lightcurves for stars with $\geq350$ days of observations and median error per visit $<0.005$ mag but without a rotation period detection (both candidate detections and non-detections are shown). The 2MASS ID and the candidate period (if identified) are shown at the top of each panel.
\label{Fig:null} }
\end{figure*}

For stars with $<350$ nights of observations or typical errors $\geq0.005$ mag, the most likely cause of a non-detection is insufficient amount or quality of data.  There persists a small number of stars for which we have gathered thousands of data points but for which a period detection has not resulted: there are $78$ stars for which the median error in one visit is $<0.005$ mag and for which observations were acquired on $\geq350$ individual nights. Of these, we detect periods in $62$. As in previous parts of this paper, we are not considering stars that are close binaries or that are strongly over-luminous.

The lightcurves for the remaining $16$ stars with high quality datasets and without period detections are shown in Fig. \ref{Fig:null}. We did not identify a single unifying feature that determined whether or not we would detect a rotation period for a star that otherwise had sufficient length and precision. The non-detections likely comprise several different categories: (1) stars with unfavorable observing conditions, (2) later-type stars, (3) long rotators with evolving spot patterns or low spot contrasts. We discuss these further below.

(1) Some stars are not favorably situated for observing with MEarth. This includes stars that have companions that are resolved, but close enough to contaminate the aperture in poor conditions. Periods close to $1$ day are also challenging to disentangle from systematics.  Our noise model includes scintillation noise and scatter in comparison stars; stars so impacted would generally be excluded by the precision cut, but these factors could contribute to systematic errors.

(2) Non-detections are biased towards later-type stars, with a drop-off in our recovery rate for $M_*<0.2$ $\msun$. The mass of all stars with and without detections are significantly different; the $p$-value of an Anderson-Darling test is $p=0.006^{+0.014}_{-0.005}$.\footnote{The Anderson-Darling k-sample test tests the null hypothesis that samples are drawn from the same distribution; our $p$-value is sufficient to reject the null hypothesis at reasonable confidence. The errors derive from a Monte Carlo simulation sampling with replacement and assuming 10\% errors on mass.} 
For the restricted subset of targets considered in this section, $81\%$ of non-detections and $60\%$ of detections have $M_*<0.2\msun$; the difference in mass distributions is not significant, but the number of stars in the samples is insufficient for the effect to be reliably seen. The bias is likely due to our diminishing ability to correct for systematics for redder stars: lower mass stars are more greatly impacted by precipitable water vapor variations, which can in particular inhibit the detection of short periods. 

(3) Most non-detections show some sort of variability, and for seven we have identified candidate long periods with periods typical of our detections. However, the photometric variability is not observed to repeat or is very low amplitude. With extended monitoring, we might happen upon an epoch in which the spot pattern is more stable or in which the contrast between spotted and unspotted photosphere is greater, and would detect a rotation period. 

A final possibility is that stars are inclined such that their poles point directly ($0\deg$) or near-directly at us; these stars would show no variability or much reduced amplitudes than if they were to be viewed edge-on. How important an effect this is depends on how spots are distributed on the stellar surface. \citet{Rebull2016a} identified rotation periods in $92\%$ of stars in the Pleiades using \ktwo, and remark that the remaining $8\%$ present unfavorable conditions (e.g. are too bright). Thus, this is not likely an important effect for rapidly rotating stars. The spot distribution of slow rotators has not been studied in as much depth. 

Considering our search for long rotation periods, (3) is the most relevant. Our hypothesis is that the non-detections shown in Fig. \ref{Fig:null} are drawn from the same population as the detections. In support of this, we detect candidate periods of similar length to our rotators for $7$ of these stars, but the variability is intermittent or quasi-periodic, and sometimes low amplitude. We suggest that the non-detections reflect the unfavorable end of the distribution of spot lifetimes, having $\tau_\mathrm{spot}\lesssim P_\mathrm{rot}$. Alternatively, this group could represent stars with a fundamentally different type of spot or dynamo behavior.

\section{Summary and discussion}

We use long-term photometric monitoring from MEarth-South to measure the rotation periods of $\numrotators$ M dwarfs in the southern hemisphere. Our rotation period search benefits from the long time-baseline and high cadence of our dataset, and consistent observing strategy. We note that we find it challenging to disentangle daily aliases in some cases; continuous monitoring from \emph{TESS} will settle the uncertainties. 

We see substantial spot evolution in some stars (see Figure \ref{Fig:gj1132}). Different modeling techniques, such as Gaussian processes \citep[e.g.][]{Angus2018} would be beneficial in these cases. Spot evolution is also important for planning exoplanet observations: as a consequence of changing surface patterns, it is optimal to have long-term photometric monitoring that coincides with transit and radial velocity measurements if we are to constrain the impact of spots and magnetic activity on these data \cite[e.g.][]{Pont2013, Rackham2018, Mallonn2018}.

The fraction of stars in which we detect a rotation period increases steadily with increasing length of the dataset. For the subset of stars with $>1200$ visits and median error $<0.005$ mag, the recovery rate is $67\pm3\%$; for comparison it was $47\pm3\%$ for northern stars subject to the same selection criteria. Considering stars observed on $\geq350$ nights and with median error $<0.005$ mag, we recover periods in $62$ of $78$ stars, and candidate periods for $7$. The spot patterns of many of these stars with detected periods seem stable for as long or longer than one rotation period ($\tau_\mathrm{spot}\gtrsim P_\mathrm{rot}$). We suggest that the non-detections are stars for which spot evolution timescales are similar to or shorter than the stellar rotation period ($\tau_\mathrm{spot}\lesssim P_\mathrm{rot}$), representing the unfavorable end of the distributions seen for long-period rotators. We do not think that there remains an unexplored population of rotators in our data. 

As a consequence, we hypothesize that we are probing the longest rotation periods typically reached by Solar Neighborhood M dwarfs; these longest periods are around $140$ days. 
The age of the local thick disk has been estimated to be $8.7\pm0.1$ Gyr  by \citet{Kilic2017} using white dwarfs within $40$ pc (which they note is younger than the canonical thick disk age at larger distances due to dynamical evolution of stars in the galaxy). The period upper limit we see could reasonably be set by a combination of this finite age in combination with Skumanich-like angular momentum loss rates \citep[see e.g.][]{Irwin2011}.

From an observational perspective, the Large Synoptic Survey Telescope \citep[LSST; see][]{Hawley2016} will provide the opportunity to probe the late-stage angular momentum evolution of M dwarfs through its precise, extensive, and long-term monitoring of M dwarfs. Though it is not expected to survey M dwarfs in clusters with ages $>5$ Gyr (the limit is younger for later-type M dwarfs), its sensitivity means that more distant and therefore sometimes older stars will be observed. If the period distribution we have detected so far results from the age of the local thick disk and M dwarfs continue to slowly spin down at ages beyond this, we should see M dwarfs with $>140$ day rotation periods with LSST.

\acknowledgments  The authors would like to acknowledge D. Kipping for encouragement in adding Proxima Centauri back into the sample after it was initially removed for being too bright. ERN is supported by an NSF Astronomy and Astrophysics Postdoctoral Fellowship under award AST-1602597, and thanks DMC-Boston for the use of a spare desk and the snack cabinet. 
NM is supported by the National Science Foundation Graduate Research Fellowship Program, under NSF grant number DGE-1745303. NM thanks the LSSTC Data Science Fellowship Program; his time as a Fellow has greatly benefited this work.
The MEarth project acknowledges funding from the National Science Foundation under grants AST-1616624, AST-0807690, AST-1109468, and AST-1004488 (Alan T. Waterman Award) and the David and Lucile Packard Foundation Fellowship for Science and Engineering. MEarth observations of GJ 1132 and LHS 1140 were supported in part through HST GO programs 14757, 14758, and 14888 which were provided by NASA through a grant from the Space Telescope Science Institute, which is operated by the Association of Universities for Research in Astronomy, Inc., under NASA contract NAS 5-26555. 
This publication was made possible through the support of a grant from the John Templeton Foundation. The opinions expressed here are those of the authors and do not necessarily reflect the views of the John Templeton Foundation. This research has made use of data products from the Two Micron All Sky Survey, which is a joint project of the University of Massachusetts and the Infrared Processing and Analysis Center / California Institute of Technology, funded by NASA and the NSF; NASA Astrophysics Data System (ADS); and the SIMBAD database and VizieR catalog access tool, at CDS, Strasbourg, France.

\clearpage


\begin{thebibliography}{}
\expandafter\ifx\csname natexlab\endcsname\relax\def\natexlab#1{#1}\fi

\bibitem[{Almenara {et~al.}(2015)Almenara, Astudillo-Defru, Bonfils, Forveille,
  Santerne, Albrecht, Barros, Bouchy, Delfosse, Demangeon, Diaz, H{\'{e}}brard,
  Mayor, Neves, Rojo, Santos, \& W{\"{u}}nsche}]{Almenara2015}
Almenara, J.~M., Astudillo-Defru, N., Bonfils, X., {et~al.} 2015, Astronomy
  {\&} Astrophysics, 581, L7

\bibitem[{Angus {et~al.}(2018)Angus, Morton, Aigrain, Foreman-Mackey, \&
  Rajpaul}]{Angus2018}
Angus, R., Morton, T., Aigrain, S., Foreman-Mackey, D., \& Rajpaul, V. 2018,
  Monthly Notices of the Royal Astronomical Society, 474, 2094

\bibitem[{Astudillo-Defru {et~al.}(2017)Astudillo-Defru, Delfosse, Bonfils,
  Forveille, Lovis, \& Rameau}]{Astudillo-Defru2017}
Astudillo-Defru, N., Delfosse, X., Bonfils, X., {et~al.} 2017, Astronomy {\&}
  Astrophysics, 600, 21

\bibitem[{Barnes(2003)}]{Barnes2003}
Barnes, S.~A. 2003, The Astrophysical Journal, 586, 464

\bibitem[{Barnes {et~al.}(2016)Barnes, Weingrill, Fritzewski, Strassmeier, \&
  Platais}]{Barnes2016}
Barnes, S.~A., Weingrill, J., Fritzewski, D., Strassmeier, K.~G., \& Platais,
  I. 2016, The Astrophysical Journal, Volume 823, Issue 1, article id. 16, 16
  pp. (2016)., 823, arXiv:1603.09179

\bibitem[{Benedict {et~al.}(1998)Benedict, McArthur, Nelan, Story, Whipple,
  Shelus, Jefferys, Hemenway, Franz, Wasserman, Duncombe, van Altena, \&
  Fredrick}]{Benedict1998}
Benedict, G.~F., McArthur, B., Nelan, E., {et~al.} 1998, The Astronomical
  Journal, 116, 429

\bibitem[{Benedict {et~al.}(2016)Benedict, Henry, Franz, McArthur, Wasserman,
  Jao, Cargile, Dieterich, Bradley, Nelan, \& Whipple}]{Benedict2016}
Benedict, G.~F., Henry, T.~J., Franz, O.~G., {et~al.} 2016, The Astronomical
  Journal, 152, 33

\bibitem[{Berta {et~al.}(2012)Berta, Irwin, Charbonneau, Burke, \&
  Falco}]{Berta2012}
Berta, Z.~K., Irwin, J., Charbonneau, D., Burke, C.~J., \& Falco, E.~E. 2012,
  The Astronomical Journal, 144, 145

\bibitem[{Berta-Thompson {et~al.}(2015)Berta-Thompson, Irwin, Charbonneau,
  Newton, Dittmann, Astudillo-Defru, Bonfils, Gillon, Jehin, Stark, Stalder,
  Bouchy, Delfosse, Forveille, Lovis, Mayor, Neves, Pepe, Santos, Udry, \&
  W{\"{u}}nsche}]{Berta-Thompson2015}
Berta-Thompson, Z.~K., Irwin, J., Charbonneau, D., {et~al.} 2015, Nature, 527,
  204

\bibitem[{Bouvier {et~al.}(2013)Bouvier, Matt, Mohanty, Scholz, Stassun, \&
  Zanni}]{Bouvier2013}
Bouvier, J., Matt, S.~P., Mohanty, S., {et~al.} 2013, in Protostars and Planets
  VI, ed. H.~Beuther, R.~S. Klessen, C.~P. Dullemond, \& T.~Henning (Tucson:
  University of Arizona Press), 433--450

\bibitem[{Boyajian {et~al.}(2012)Boyajian, von Braun, van Belle, McAlister, ten
  Brummelaar, Kane, Muirhead, Jones, White, Schaefer, Ciardi, Henry,
  L{\'{o}}pez-Morales, Ridgway, Gies, Jao, Rojas-Ayala, Parks, Sturmann,
  Sturmann, Turner, Farrington, Goldfinger, \& Berger}]{Boyajian2012}
Boyajian, T.~S., von Braun, K., van Belle, G., {et~al.} 2012, The Astrophysical
  Journal, 757, 112

\bibitem[{Browning {et~al.}(2010)Browning, Basri, Marcy, West, \&
  Zhang}]{Browning2010}
Browning, M.~K., Basri, G., Marcy, G.~W., West, A.~A., \& Zhang, J. 2010, The
  Astronomical Journal, 139, 504

\bibitem[{Cloutier {et~al.}(2016)Cloutier, Doyon, Menou, Delfosse, Dumusque, \&
  Artigau}]{Cloutier2016}
Cloutier, R., Doyon, R., Menou, K., {et~al.} 2016, The Astronomical Journal,
  153, 9

\bibitem[{Cohen {et~al.}(2018)Cohen, Glocer, Garraffo, Drake, \&
  Bell}]{Cohen2018}
Cohen, O., Glocer, A., Garraffo, C., Drake, J.~J., \& Bell, J.~M. 2018, The
  Astrophysical Journal, 856, L11

\bibitem[{Covey {et~al.}(2016)Covey, Ag{\"{u}}eros, Law, Liu, Ahmadi, Laher,
  Levitan, Sesar, \& Surface}]{Covey2016}
Covey, K.~R., Ag{\"{u}}eros, M.~A., Law, N.~M., {et~al.} 2016, The
  Astrophysical Journal, 822, 27

\bibitem[{Damasso {et~al.}(2018)Damasso, Bonomo, Astudillo-Defru, Bonfils,
  Malavolta, Sozzetti, Lopez, Zeng, Haywood, Irwin, Mortier, Vanderburg,
  Maldonado, Lanza, Affer, Almenara, Benatti, Biazzo, Bignamini, Borsa, Bouchy,
  Buchhave, Cameron, Carleo, Charbonneau, Claudi, Cosentino, Covino, Delfosse,
  Desidera, {Di Fabrizio}, Dressing, Esposito, Fares, Figueira, Fiorenzano,
  Forveille, Giacobbe, Gonz{\'{a}}lez-{\'{A}}lvarez, Gratton, Harutyunyan,
  Johnson, Latham, Leto, Lopez-Morales, Lovis, Maggio, Mancini, Masiero, Mayor,
  Micela, Molinari, Motalebi, Murgas, Nascimbeni, Pagano, Pepe, Phillips,
  Piotto, Poretti, Rainer, Rice, Santos, Sasselov, Scandariato,
  S{\'{e}}gransan, Smareglia, Udry, Watson, \& W{\"{u}}nsche}]{Damasso2018}
Damasso, M., Bonomo, A.~S., Astudillo-Defru, N., {et~al.} 2018, eprint
  arXiv:1802.08320

\bibitem[{Delfosse {et~al.}(2000)Delfosse, Forveille, S{\'{e}}gransan, Beuzit,
  Udry, Perrier, \& Mayor}]{Delfosse2000}
Delfosse, X., Forveille, T., S{\'{e}}gransan, D., {et~al.} 2000, Astronomy and
  Astrophysics, 364, 217

\bibitem[{Dittmann {et~al.}(2017)Dittmann, Irwin, Charbonneau, Bonfils,
  Astudillo-Defru, Haywood, Berta-Thompson, Newton, Rodriguez, Winters, Tan,
  Almenara, Bouchy, Delfosse, Forveille, Lovis, Murgas, Pepe, Santos, Udry,
  W{\"{u}}nsche, Esquerdo, Latham, \& Dressing}]{Dittmann2017}
Dittmann, J.~A., Irwin, J.~M., Charbonneau, D., {et~al.} 2017, Nature, 544, 333

\bibitem[{Douglas {et~al.}(2016)Douglas, Ag{\"{u}}eros, Covey, Cargile,
  Barclay, Cody, Howell, \& Kopytova}]{Douglas2016}
Douglas, S.~T., Ag{\"{u}}eros, M.~A., Covey, K.~R., {et~al.} 2016, The
  Astrophysical Journal, 822, 47

\bibitem[{Douglas {et~al.}(2017)Douglas, Ag{\"{u}}eros, Covey, \&
  Kraus}]{Douglas2017}
Douglas, S.~T., Ag{\"{u}}eros, M.~A., Covey, K.~R., \& Kraus, A. 2017, The
  Astrophysical Journal, 842, 83

\bibitem[{Garraffo {et~al.}(2017)Garraffo, Drake, Cohen, Alvarado-G{\'{o}}mez,
  \& Moschou}]{Garraffo2017}
Garraffo, C., Drake, J.~J., Cohen, O., Alvarado-G{\'{o}}mez, J.~D., \& Moschou,
  S.~P. 2017, The Astrophysical Journal, 843, L33

\bibitem[{Gliese \& Jahrei{\ss}(1991)}]{Gliese1991}
Gliese, W., \& Jahrei{\ss}, H. 1991, On: The Astronomical Data Center CD-ROM:
  Selected Astronomical Catalogs

\bibitem[{Hartman {et~al.}(2010)Hartman, Bakos, Kov{\'{a}}cs, \&
  Noyes}]{Hartman2010}
Hartman, J.~D., Bakos, G.~{\'{A}}., Kov{\'{a}}cs, G., \& Noyes, R.~W. 2010,
  Monthly Notices of the Royal Astronomical Society, 408, 475

\bibitem[{Hartman {et~al.}(2011)Hartman, Bakos, Noyes, Sipőcz, Kov{\'{a}}cs,
  Mazeh, Shporer, \& P{\'{a}}l}]{Hartman2011}
Hartman, J.~D., Bakos, G.~{\'{A}}., Noyes, R.~W., {et~al.} 2011, The
  Astronomical Journal, 141, 166

\bibitem[{Hawley {et~al.}(2016)Hawley, Angus, Buzasi, Davenport, Giampapa,
  Kashyap, \& Meibom}]{Hawley2016}
Hawley, S.~L., Angus, R., Buzasi, D., {et~al.} 2016, eprint arXiv:1607.04302

\bibitem[{Hawley {et~al.}(1996)Hawley, Gizis, \& Reid}]{Hawley1996}
Hawley, S.~L., Gizis, J.~E., \& Reid, I.~N. 1996, The Astronomical Journal,
  112, 2799

\bibitem[{Henry {et~al.}(1994)Henry, Kirkpatrick, \& Simons}]{Henry1994}
Henry, T.~J., Kirkpatrick, J.~D., \& Simons, D.~A. 1994, The Astronomical
  Journal, 108, 1437

\bibitem[{Henry {et~al.}(2004)Henry, Subasavage, Brown, Beaulieu, Jao, \&
  Hambly}]{Henry2004}
Henry, T.~J., Subasavage, J.~P., Brown, M.~A., {et~al.} 2004, The Astronomical
  Journal, Volume 128, Issue 5, pp. 2460-2473., 128, 2460

\bibitem[{Irwin {et~al.}(2006)Irwin, Aigrain, Hodgkin, Irwin, Bouvier, Clarke,
  Hebb, \& Moraux}]{Irwin2006}
Irwin, J., Aigrain, S., Hodgkin, S., {et~al.} 2006, Monthly Notices of the
  Royal Astronomical Society, 370, 954

\bibitem[{Irwin {et~al.}(2011)Irwin, Berta, Burke, Charbonneau, Nutzman, West,
  \& Falco}]{Irwin2011}
Irwin, J., Berta, Z.~K., Burke, C.~J., {et~al.} 2011, The Astrophysical
  Journal, 727, 56

\bibitem[{Irwin {et~al.}(2007)Irwin, Hodgkin, Aigrain, Hebb, Bouvier, Clarke,
  Moraux, \& Bramich}]{Irwin2007}
Irwin, J., Hodgkin, S., Aigrain, S., {et~al.} 2007, Monthly Notices of the
  Royal Astronomical Society, 377, 741

\bibitem[{Irwin {et~al.}(2015)Irwin, Berta-Thompson, Charbonneau, Dittmann,
  Falco, Newton, \& Nutzman}]{Irwin2015}
Irwin, J.~M., Berta-Thompson, Z.~K., Charbonneau, D., {et~al.} 2015, in 18th
  Cambridge Workshop on Cool Stars, Stellar Systems, and the Sun, ed. G.~van
  Belle \& H.~Harris., 767--772

\bibitem[{Kado-Fong {et~al.}(2016)Kado-Fong, Williams, Mann, Berger, Burgett,
  Chambers, Huber, Kaiser, Kudritzki, Magnier, Wainscoat, \&
  Waters}]{Kado-Fong2016}
Kado-Fong, E., Williams, P. K.~G., Mann, A.~W., {et~al.} 2016, The
  Astrophysical Journal, 833, 1

\bibitem[{Kilic {et~al.}(2017)Kilic, Munn, Harris, von Hippel, Liebert,
  Williams, Jeffery, \& DeGennaro}]{Kilic2017}
Kilic, M., Munn, J.~A., Harris, H.~C., {et~al.} 2017, The Astrophysical
  Journal, 837, 162

\bibitem[{Kiraga \& Stepien(2007)}]{Kiraga2007}
Kiraga, M., \& Stepien, K. 2007, Acta Astronomica, 57, 149

\bibitem[{Kirkpatrick {et~al.}(1991)Kirkpatrick, Henry, \&
  McCarthy}]{Kirkpatrick1991}
Kirkpatrick, J.~D., Henry, T.~J., \& McCarthy, D.~W. 1991, The Astrophysical
  Journal Supplement Series, 77, 417

\bibitem[{Kraft(1967)}]{Kraft1967}
Kraft, R.~P. 1967, The Astrophysical Journal, 150, 551

\bibitem[{Lammer {et~al.}(2007)Lammer, Lichtenegger, Kulikov, Grie{\ss}meier,
  Terada, Erkaev, Biernat, Khodachenko, Ribas, Penz, \& Selsis}]{Lammer2007}
Lammer, H., Lichtenegger, H.~I., Kulikov, Y.~N., {et~al.} 2007, Astrobiology,
  7, 185

\bibitem[{L{\'{e}}pine(2005)}]{Lepine2005}
L{\'{e}}pine, S. 2005, The Astronomical Journal, 130, 1680

\bibitem[{L{\'{e}}pine \& Gaidos(2011)}]{Lepine2011}
L{\'{e}}pine, S., \& Gaidos, E. 2011, The Astronomical Journal, 142, 138

\bibitem[{L{\'{e}}pine \& Shara(2005)}]{Lepine2005a}
L{\'{e}}pine, S., \& Shara, M.~M. 2005, The Astronomical Journal, 129, 1483

\bibitem[{Luger \& Barnes(2015)}]{Luger2015}
Luger, R., \& Barnes, R. 2015, Astrobiology, 15, 119

\bibitem[{Mallonn {et~al.}(2018)Mallonn, Herrero, Juvan, von Essen, Rosich,
  Ribas, Granzer, Alexoudi, \& Strassmeier}]{Mallonn2018}
Mallonn, M., Herrero, E., Juvan, I.~G., {et~al.} 2018, eprint arXiv:1803.05677

\bibitem[{Matt {et~al.}(2015)Matt, Brun, Baraffe, Bouvier, \&
  Chabrier}]{Matt2015}
Matt, S.~P., Brun, A.~S., Baraffe, I., Bouvier, J., \& Chabrier, G. 2015, The
  Astrophysical Journal, 799, L23

\bibitem[{McQuillan {et~al.}(2013)McQuillan, Aigrain, \& Mazeh}]{McQuillan2013}
McQuillan, A., Aigrain, S., \& Mazeh, T. 2013, Monthly Notices of the Royal
  Astronomical Society, 432, 1203

\bibitem[{Newton {et~al.}(2017)Newton, Irwin, Charbonneau, Berlind, Calkins, \&
  Mink}]{Newton2017}
Newton, E.~R., Irwin, J., Charbonneau, D., {et~al.} 2017, The Astrophysical
  Journal, 834, 85

\bibitem[{Newton {et~al.}(2016{\natexlab{a}})Newton, Irwin, Charbonneau,
  Berta-Thompson, \& Dittmann}]{Newton2016a}
Newton, E.~R., Irwin, J., Charbonneau, D., Berta-Thompson, Z.~K., \& Dittmann,
  J.~A. 2016{\natexlab{a}}, The Astrophysical Journal, 821, L19

\bibitem[{Newton {et~al.}(2016{\natexlab{b}})Newton, Irwin, Charbonneau,
  Berta-Thompson, Dittmann, \& West}]{Newton2016}
Newton, E.~R., Irwin, J., Charbonneau, D., {et~al.} 2016{\natexlab{b}}, The
  Astrophysical Journal, 821, 93

\bibitem[{Nutzman \& Charbonneau(2008)}]{Nutzman2008}
Nutzman, P., \& Charbonneau, D. 2008, Publications of the Astronomical Society
  of the Pacific, 120, 317

\bibitem[{Pojmanski(2002)}]{Pojmanski2002}
Pojmanski, G. 2002, Acta Astronomica, 52, 397

\bibitem[{Pont {et~al.}(2013)Pont, Sing, Gibson, Aigrain, Henry, \&
  Husnoo}]{Pont2013}
Pont, F., Sing, D.~K., Gibson, N.~P., {et~al.} 2013, Monthly Notices of the
  Royal Astronomical Society, 432, 2917

\bibitem[{Rackham {et~al.}(2018)Rackham, Apai, \& Giampapa}]{Rackham2018}
Rackham, B.~V., Apai, D., \& Giampapa, M.~S. 2018, eprint arXiv:1711.05691

\bibitem[{Rebull {et~al.}(2016)Rebull, Stauffer, Bouvier, Cody, Hillenbrand,
  Soderblom, Valenti, Barrado, Bouy, Ciardi, Pinsonneault, Stassun, Micela,
  Aigrain, Vrba, Somers, Christiansen, Gillen, \& Cameron}]{Rebull2016a}
Rebull, L.~M., Stauffer, J.~R., Bouvier, J., {et~al.} 2016, The Astronomical
  Journal, 152, 113

\bibitem[{Reid {et~al.}(1995)Reid, Hawley, \& Gizis}]{Reid1995}
Reid, I.~N., Hawley, S.~L., \& Gizis, J.~E. 1995, The Astronomical Journal,
  110, 1838

\bibitem[{Reiners \& Mohanty(2012)}]{Reiners2012b}
Reiners, A., \& Mohanty, S. 2012, The Astrophysical Journal, 746, 43

\bibitem[{Ricker {et~al.}(2014)Ricker, Winn, Vanderspek, Latham, Bakos, Bean,
  Berta-Thompson, Brown, Buchhave, Butler, Butler, Chaplin, Charbonneau,
  Christensen-Dalsgaard, Clampin, Deming, Doty, {De Lee}, Dressing, Dunham,
  Endl, Fressin, Ge, Henning, Holman, Howard, Ida, Jenkins, Jernigan, Johnson,
  Kaltenegger, Kawai, Kjeldsen, Laughlin, Levine, Lin, Lissauer, MacQueen,
  Marcy, McCullough, Morton, Narita, Paegert, Palle, Pepe, Pepper, Quirrenbach,
  Rinehart, Sasselov, Sato, Seager, Sozzetti, Stassun, Sullivan, Szentgyorgyi,
  Torres, Udry, \& Villasenor}]{Ricker2014}
Ricker, G.~R., Winn, J.~N., Vanderspek, R., {et~al.} 2014, Journal of
  Astronomical Telescopes, Instruments, and Systems, 1, 014003

\bibitem[{Robertson {et~al.}(2014)Robertson, Mahadevan, Endl, \&
  Roy}]{Robertson2014}
Robertson, P., Mahadevan, S., Endl, M., \& Roy, A. 2014, Science, 345, 440

\bibitem[{Schatzman(1962)}]{Schatzman1962}
Schatzman, E. 1962, Annales d'Astrophysique, 25, 18

\bibitem[{Seabold \& Perktold(2010)}]{Seabold2010}
Seabold, S., \& Perktold, J. 2010, in Proceedings of the 9th Python in Science
  Conference, ed. S.~van~der Walt \& M.~Jarrod, 57--61

\bibitem[{Smith {et~al.}(2012)Smith, Stumpe, {Van Cleve}, Jenkins, Barclay,
  Fanelli, Girouard, Kolodziejczak, McCauliff, Morris, \& Twicken}]{Smith2012}
Smith, J.~C., Stumpe, M.~C., {Van Cleve}, J.~E., {et~al.} 2012, Publications of
  the Astronomical Society of the Pacific, 124, 1000

\bibitem[{Stumpe {et~al.}(2012)Stumpe, Smith, {Van Cleve}, Twicken, Barclay,
  Fanelli, Girouard, Jenkins, Kolodziejczak, McCauliff, \& Morris}]{Stumpe2012}
Stumpe, M.~C., Smith, J.~C., {Van Cleve}, J.~E., {et~al.} 2012, Publications of
  the Astronomical Society of the Pacific, 124, 985

\bibitem[{Sullivan {et~al.}(2015)Sullivan, Winn, Berta-Thompson, Charbonneau,
  Deming, Dressing, Latham, Levine, McCullough, Morton, Ricker, Vanderspek, \&
  Woods}]{Sullivan2015}
Sullivan, P.~W., Winn, J.~N., Berta-Thompson, Z.~K., {et~al.} 2015, The
  Astrophysical Journal, 809, 77

\bibitem[{{The Astropy Collaboration} {et~al.}(2018){The Astropy
  Collaboration}, Price-Whelan, Sipőcz, G{\"{u}}nther, Lim, Crawford, Conseil,
  Shupe, Craig, Dencheva, Ginsburg, VanderPlas, Bradley,
  P{\'{e}}rez-Su{\'{a}}rez, de~Val-Borro, Aldcroft, Cruz, Robitaille, Tollerud,
  Ardelean, Babej, Bachetti, Bakanov, Bamford, Barentsen, Barmby, Baumbach,
  Berry, Biscani, Boquien, Bostroem, Bouma, Brammer, Bray, Breytenbach,
  Buddelmeijer, Burke, Calderone, Rodr{\'{i}}guez, Cara, Cardoso, Cheedella,
  Copin, Crichton, D{\'{A}}vella, Deil, Depagne, Dietrich, Donath, Droettboom,
  Earl, Erben, Fabbro, Ferreira, Finethy, Fox, Garrison, Gibbons, Goldstein,
  Gommers, Greco, Greenfield, Groener, Grollier, Hagen, Hirst, Homeier, Horton,
  Hosseinzadeh, Hu, Hunkeler, Ivezi{\'{c}}, Jain, Jenness, Kanarek, Kendrew,
  Kern, Kerzendorf, Khvalko, King, Kirkby, Kulkarni, Kumar, Lee, Lenz,
  Littlefair, Ma, Macleod, Mastropietro, McCully, Montagnac, Morris, Mueller,
  Mumford, Muna, Murphy, Nelson, Nguyen, Ninan, N{\"{o}}the, Ogaz, Oh, Parejko,
  Parley, Pascual, Patil, Patil, Plunkett, Prochaska, Rastogi, Janga, Sabater,
  Sakurikar, Seifert, Sherbert, Sherwood-Taylor, Shih, Sick, Silbiger,
  Singanamalla, Singer, Sladen, Sooley, Sornarajah, Streicher, Teuben, Thomas,
  Tremblay, Turner, Terr{\'{o}}n, van Kerkwijk, de~la Vega, Watkins, Weaver,
  Whitmore, Woillez, \& Zabalza}]{TheAstropyCollaboration2018}
{The Astropy Collaboration}, T.~A., Price-Whelan, A.~M., Sipőcz, B.~M.,
  {et~al.} 2018, eprint arXiv:1801.02634

\bibitem[{Vanderburg \& Johnson(2014)}]{Vanderburg2014}
Vanderburg, A., \& Johnson, J.~A. 2014, Publications of the Astronomical
  Society of Pacific, 126, 948

\bibitem[{Vanderburg {et~al.}(2016)Vanderburg, Plavchan, Johnson, Ciardi,
  Swift, \& Kane}]{Vanderburg2016}
Vanderburg, A., Plavchan, P., Johnson, J.~A., {et~al.} 2016, Monthly Notices of
  the Royal Astronomical Society, 459, 3565

\bibitem[{Wargelin {et~al.}(2017)Wargelin, Saar, Pojma{\'{n}}ski, Drake, \&
  Kashyap}]{Wargelin2017}
Wargelin, B.~J., Saar, S.~H., Pojma{\'{n}}ski, G., Drake, J.~J., \& Kashyap,
  V.~L. 2017, Monthly Notices of the Royal Astronomical Society, 464, 3281

\bibitem[{Weinberger {et~al.}(2016)Weinberger, Boss, Keiser,
  Anglada-Escud{\'{e}}, Thompson, \& Burley}]{Weinberger2016}
Weinberger, A.~J., Boss, A.~P., Keiser, S.~A., {et~al.} 2016, The Astronomical
  Journal, 152, 24

\bibitem[{Winters {et~al.}(2015)Winters, Henry, Lurie, Hambly, Jao, Bartlett,
  Boyd, Dieterich, Finch, Hosey, Ianna, Riedel, Slatten, \&
  Subasavage}]{Winters2015}
Winters, J.~G., Henry, T.~J., Lurie, J.~C., {et~al.} 2015, The Astronomical
  Journal, 149, 5

\end{thebibliography}
\end{document}